  \documentclass[prd,aps,twocolumn,a4paper,nofootinbib]{revtex4-1}

\newif\ifusesec
\usesectrue % or
\usepackage{graphicx,psfrag}
\usepackage{mathrsfs}
\usepackage{amsmath,amsfonts,amssymb}
\usepackage{multirow}
\usepackage{comment}

\DeclareSymbolFontAlphabet{\mathrsfs}{rsfs}
\DeclareMathAlphabet\mathbfcal{OMS}{cmsy}{b}{n}

\newcommand{\be}{\begin{equation}}
\newcommand{\ee}{\end{equation}}
\newcommand{\bea}{\begin{eqnarray}}
\newcommand{\eea}{\end{eqnarray}}
\newcommand{\bel}{\begin{align}}
\newcommand{\eel}{\end{align}}

\def\GMc2{G M_{\odot} c^{-2}}

\def\J{{\cal J}}
\def\E{{\cal E}}
\def\lm{{\ell m}}

\usepackage{color}

\definecolor{cyan}{rgb}{0,0.9,0.9}
\definecolor{orange}{rgb}{0.9,0.5,0}
\definecolor{magenta}{rgb}{1,0,1}
\definecolor{purple}{rgb}{0.8,0.4,0.8}
\definecolor{gray}{rgb}{0.8242,0.8242,0.8242}

\begin{document}

\title{Tidal effects in binary neutron star coalescence}

\author{Sebastiano \surname{Bernuzzi}$^1$}
\author{Alessandro \surname{Nagar}$^2$}
\author{Marcus \surname{Thierfelder}$^1$}
\author{Bernd \surname{Br\"ugmann}$^1$}

\affiliation{$^1$Theoretical Physics Institute, University of Jena, 07743 Jena, Germany}
\affiliation{$^2$Institut des Hautes Etudes Scientifiques, 91440 Bures-sur-Yvette, France}

\date{\today}

\begin{abstract}
  We compare dynamics and waveforms from binary neutron star
  coalescence as computed by new long-term ($\sim 10 $ orbits) 
  numerical relativity simulations and by the tidal effective-one-body (EOB) 
  model including analytical tidal corrections up to second post-Newtonian order (2PN).
  The current analytic knowledge encoded in the tidal EOB model is found 
  to be sufficient to reproduce the numerical data up to contact and within 
  their uncertainties. Remarkably, no calibration of any tidal EOB free parameters 
  is required, beside those already fitted to binary black holes data.
  The inclusion of 2PN tidal corrections minimizes the differences with 
  the numerical data, but it is not possible to significantly distinguish 
  them from the leading-order tidal contribution. The presence of a relevant 
  amplification of tidal effects is likely to be excluded, although it 
  can appear as a consequence of numerical inaccuracies.
  We conclude that the tidally-completed effective-one-body model
  provides nowadays the most advanced and accurate tool for modelling
  gravitational waveforms from binary neutron star inspiral up to contact.
  This work also points out the importance of extensive tests to assess 
  the uncertainties of the numerical data, and the potential need of new 
  numerical strategies to perform accurate simulations.   
\end{abstract}


\maketitle

\ifusesec
\section{Introduction}
\else
\paragraph*{Introduction}
\fi
\label{sec:intro}

Gravitational waves (GWs) emitted by binary neutron star (BNS)
inspiral and coalescence will be detectable by advanced LIGO-VIRGO
detectors. The tidal signature in such waves is (mainly) proportional
to the {\it tidal polarizability parameter} $\mu_2$ that yields the
ratio between the tidally induced quadrupole moment and the
companion's perturbing tidal gradient.  The tidal parameter $\mu_2$
depends on the neutron star equation of state (EOS) and it is related
to the relativistic generalization of the Newtonian, dimensionless,
Love
number~\cite{Damour:1983tz,Hinderer:2007mb,Damour:2009kr,Binnington:2009bb,Hinderer:2009ca}
$k_2$ as $\mu_2=2/(3G)k_2R^5$, where $R$ is the star radius and $G$
the Newton constant. The late-inspiral part of the GW signal, where
tidal effects are stronger, can be used to measure the tidal Love
number and thus to extract information about the nuclear EOS.  A
recent study~\cite{Damour:2012yf} of the measurability of $G\mu_2$,
based on the tidal extension~\cite{Damour:2009wj} of the
effective-one-body (EOB)
model~\cite{Buonanno:1998gg,Buonanno:2000ef,Damour:2000we,Damour:2008gu,Damour:2009kr},
has shown that from a detection of GWs up to merger all normal matter
content ($npe\mu$) EOS with maximum mass $\gtrsim1.97~M_{\odot}$ can 
be distinguished at $95\%$ confidence with signal-to-noise (SNR) ratio 
$16$ and for any physical
mass ratio~\footnote{ 
  On the contrary, if only the early inspiral
  waveform is considered, i.e.~only GW frequencies $\lesssim450$~Hz (for
  a BNS with individual masses $1.4~M_{\odot}$) it is not possible to
  measure $G\mu_2$ with sufficient accuracy to discriminate among
  different EOS~\cite{Hinderer:2009ca}.  }. 

Accurate theoretical modelling of GWs from BNS coalescence is a
challenging task. High post-Newtonian (PN) tidal corrections and
resummation techniques are needed to push the validity of the (semi)
analytical models up to contact. Next-to-leading order (NLO,
fractional 1PN accuracy)~\cite{Damour:2009wj} (then confirmed
in~\cite{Vines:2010ca}) and next-to-next-to-leading-order (NNLO,
fractional 2PN accuracy) relativistic corrections to the tidal
interaction energy have been computed recently using
effective-field-theory techniques~\cite{Bini:2012gu}.  Fractional 1PN
tidal corrections to the waveform were also obtained
in~\cite{Vines:2011ud}.  The high-PN tidal corrections effectively
amplify the magnitude of leading-order tidal effects, and are now incorporated in
the tidal EOB model~\cite{Damour:2009vw,Damour:2009wj,Damour:2012yf},
which is currently the most sophisticated analytical tool available to
model the dynamics and waveforms of neutron star (or even mixed)
binaries up to contact.

Numerical relativity (NR) simulations are the fundamental tool to
compute the dynamics and waveform of the last few orbits of a 
coalescing BNS system. NR data can be used to
calibrate yet uncalculated higher-order tidal effects and to tests 
the reliability of the analytical models. To date, however, 
only few works have explored this important
problem~\cite{Read:2009yp,Baiotti:2010xh,Bernuzzi:2011aq}.  

A first comparison~\cite{Read:2009yp} between waveforms from
three-orbits NR simulations and the standard, point-mass,
Taylor-T4 PN approximant, pointed out that the dephasing
accumulated during the last orbits up to merger can be observed and
used to constrain the EOS. As discussed there, a major limitation of
that work was probably given by the length of the NR data  available
at the time.  

Long-term (nine and eleven orbits) BNS numerical simulations were
presented in Refs.~\cite{Baiotti:2010xh,Baiotti:2011am}, and compared
there with the prediction of the tidal EOB model. By performing a
gauge-invariant and frequency-based analysis of the phasing,
it was found that the tidal interaction predicted by the
numerical simulation is important even in the early part of the signal.
To model it analytically, it was necessary to introduce  effective 
fractional 2PN tidal corrections to the EOB model, yielding an 
amplification of the analytically predicted tidal effects~\cite{Damour:2009wj}, 
via the free parameter $\bar{\alpha}_2$; the simulations constrained 
it~\footnote{ 
  A similar conclusion, $\bar{\alpha}_2\sim 40$, was also reached 
in~\cite{Damour:2009wj} using non-conformally-flat, 
NR stationary BNS sequencies.}
in the range $40\lesssim\bar{\alpha}_2\lesssim 100$.
That analysis also pointed out that a more detailed analysis of
finite-resolution uncertainties on long-inspiral BNS waveforms was
needed to correctly estimate the magnitude of tidal effects.

Recently, this important task was undertaken
in~\cite{Bernuzzi:2011aq}, that presented the first comprehensive
analysis of the uncertainties on the waveforms due to truncation and
finite extraction error in a nine-orbit BNS simulation. In the same
work, NR waveforms were compared to the tidal T4
approximant~\cite{Hinderer:2009ca} including also NLO tidal
corrections~\cite{Damour:2009wj,Vines:2010ca,Vines:2011ud}.
Significant effects on the phasing due to high-order tidal effects
were observed in the ``best data'' (extrapolated from various
resolutions), although a more conservative error estimate did not
allow to distinguish higher-order tidal effects.

Motivated by these works and by the last analytical results
of~\cite{Bini:2012gu}, we address in this paper the following
question:
Is the current analytical knowledge necessary and sufficient to
reproduce NR data within their uncertainties?

We present results about new NR long-term simulations and their
comparison with the up-to-date EOB model
~\cite{Damour:2009vw,Damour:2009wj} which includes the NNLO tidal 
corrections of~\cite{Bini:2012gu}. Two different sets of simulations of the same
initial data are considered and some difficulties in obtaining NR data
of sufficient accuracy are pointed out.  We experimentally estimate
the contact frequency of the binary, a fundamental information for a
comparison with analytical models.  NR data are then compared to the EOB
model by carefully taking into account their uncertainties. We make
use of the gauge-invariant relation between the (reduced) binding
energy $E$ and the (reduced) angular momentum $j$ of the system in
order to analyse the dynamics of the binary~\cite{Damour:2011fu}. The
phasing of the waveforms is studied both in the time domain and by
means of a gauge-invariant and frequency-based approach which does 
not require to fix any relative (phase and time) alignment between
the waveforms~\cite{Baiotti:2011am}.

\ifusesec
The paper is organized as follows. 
In Sec.~\ref{sec:anal} the tidal EOB model used here is reviewed.
In Sec.~\ref{sec:nr} the numerical simulations are presented.
The EOB/NR comparison is discussed in Sec.~\ref{sec:res}.
Concluding remarks are in Sec.~\ref{sec:conc}. 
\else

\fi
We use units $G=c=M_\odot=1$, unless otherwise stated.

\ifusesec
\section{Tidal 2PN EOB model}
\else
\paragraph*{Tidal 2PN EOB model}
\fi
\label{sec:anal}

The tidal extension of the EOB model of the binary dynamics has been defined in Ref.~\cite{Damour:2009wj}
and then improved in Ref.~\cite{Bini:2012gu} and~\cite{Damour:2012yf} to include fractional 
2PN corrections in the tidal part of the EOB potential $A(r)$. In particular, we address
the reader to Appendix~A of Ref.~\cite{Damour:2012yf} for a collection of ready-to-use 
formulas that define the EOB dynamics and waveforms including fractional 2PN tidal effects.
Here we only summarize the main points.

The EOB radial potential has the form
\be
A(u) = A^0(u)+A^{\rm tidal}(u),
\ee
where $u\equiv 1/r=GM/(c^2 r_{\rm AB})$ is the Newtonian potential, $M=M_A+M_B$ the total mass, 
$r_{\rm AB}$ the relative separation, $A^0(u)$ denotes the point-mass potential 
and $A^{\rm tidal}(u)$ is the supplementary tidal contribution of the form
\be
\label{eq:atidal}
A^{\rm tidal}(u) = \sum_{\ell =2}^{4}-\kappa_\ell^T u^{2\ell+2}\hat{A}_{\rm tidal}^{(\ell)}(u).
\ee
The point-mass potential is defined using the usual Pad\'e resummation of the 5PN 
Taylor expansion of the $A$ function with the 4PN and 5PN EOB parameters $(a_5,a_6)$, 
$A^0(u)=P^1_5[1-2u + 2\nu u^3 + a_4\nu u^4 + a_5\nu u^5 + a_6 \nu u^6]$, where
$a_4=94/3 - (41/32)\pi^2$, $\nu=M_A M_B/M^2$, and $P^n_m$ denotes an $(n,m)$ 
Pad\'e approximant. Following the finding of Ref.~\cite{Damour:2009kr} (then substantially 
confirmed by Ref.~\cite{Pan:2011gk}) we fix the free EOB parameters to the 
the values $a_5=-6.37$ and $a_6=+50$ which lie within the extended region in
the $(a_5,a_6)$ plane yielding a good fit of the binary black hole equal-mass
simulations.

In the tidal contribution, Eq.~\eqref{eq:atidal}, the terms $\kappa_\ell^T u^{2\ell +2}$ represent 
the leading-order (LO) tidal interaction, while the additional factor $\hat{A}_{\rm tidal}^{(\ell)}(u)$ 
takes into account the effect of distance dependent, higher-order relativistic contributions 
to the dynamical tidal interactions:
1PN (first-order in $u$, or next-to-leading order, NLO), 2PN (of order $u^2$, or next-to-next-to-leading
order, NNLO), etc. The dimensionless EOB tidal parameter $\kappa_\ell^T$ is related to the tidal
polarizability coefficients $G\mu_\ell^{A,B}$ of each neutron star as
\be
\kappa_\ell^T \equiv \kappa_\ell^A + \kappa_\ell^B
\ee
where
\be
\kappa_\ell^A \equiv (2\ell-1)!!\dfrac{M_B}{M_A}\dfrac{G\mu_\ell^A}{(GM/c^2)^{2\ell+1}}.
\ee
Here we take advantage of the new analytical results of Ref.~\cite{Bini:2012gu} 
and we use the newly computed expressions of $\hat{A}^{(\ell)}_{\rm tidal}(u)$ for $\ell=2$ 
and $\ell=3$ at NNLO accuracy. Focusing on the most relevant equal-mass case ($\nu=1/4$), 
the relativistic correction to the tidal potential reads
\be
\hat{A}_{\rm tidal}^{(\ell)}(u)=1 + \alpha_1^{(\ell)}u + \alpha_2^{(\ell)}u^2,
\ee
where the coefficients $\alpha_{1,2}^{(\ell)}$ are, in this particular case, pure numbers. 
Specializing to the equal-mass case Eqs.~(6.9)-(6.10) and Eqs.~(6.21) and~(6.22) of
Ref.~\cite{Bini:2012gu} we obtain
\begin{align}
\label{eqa}
\hat{A}_{\rm tidal}^{(2)}(u) &= 1 + \dfrac{5}{4} u + \dfrac{85}{14} u^2,\\
\label{eqb}
\hat{A}_{\rm tidal}^{(3)}(u) &= 1 + \dfrac{7}{4} u + \dfrac{257}{48} u^2, \\
\label{eqc}
\hat{A}_{\rm tidal}^{(4)}(u) &= 1,
\end{align}
where we indicated explicitly the absence of (yet uncomputed)
higher-order corrections to the $\ell=4$ relativistic
contribution. The EOB waveform and radiation reaction is computed as
in Refs.~\cite{Baiotti:2011am} and takes explicitly into account the
1PN tidal corrections of~\cite{Vines:2011ud} (see Appendix~A
of~\cite{Damour:2012yf} for the precise definition of the EOB waveform
with tidal corrections).  

Equations~\eqref{eqa}-\eqref{eqc} define the most advanced tidal EOB model based on
analytical information only. In Ref.~\cite{Baiotti:2011am} a slightly simplified 
representation of the functions $\hat{A}_{\rm tidal}^{(\ell)}$ was used. Since at 
the time the NNLO calculation was not completed yet, and only
$\alpha_1^{(2)}=5/4$ was known, one was using the following NNLO
effective expression for the relativistic tidal corrections 
\be
\label{eq:Aeff}
\hat{A}_{\rm tidal}^{(\ell)}=1+\bar{\alpha}_1 u +\bar{\alpha}_2 u^2,
\ee 
with $\bar{\alpha}_1\equiv \alpha_1^{(2)}=5/4$ fixed to be the {\it
  same} for $\ell=2,3,4$ 
and $\bar{\alpha}_2$ taken as a {\it free} effective parameter (for all $\ell$'s) 
to be fitted for by comparison with NR simulations.
Although in the following we shall mainly focus on the purely
analytical 2PN tidal EOB model defined by
Eqs.~\eqref{eqa}-\eqref{eqb}, we shall also contrast some predictions
of the {\it effective} 2PN EOB model given by Eq.~\ref{eq:Aeff} with
the numerical data. 

The last important concept we want to remember is the definition of
{\it contact} between the two stars. This quantity is important
because the analytical model ceases, in principle, to be valid after
this moment. Such a formal contact moment was introduced in Eqs.~(72)
and (77) of~\cite{Damour:2009wj} by the condition that the EOB radial
separation $r_{AB}$ becomes equal to the sum of the tidally deformed
radii of the two stars, namely  
\be
\label{eq:eob_contact}
r_{AB}^{\rm contact}=\left(1+h_2^A \epsilon_A(r_{AB}^{\rm
  contact})\right)R_A + {A\leftrightarrow B}, 
\ee
where $\epsilon_A(r_{AB})=M_BR_A^3/(r_{AB}^3 M)$ is the dimensionless
parameter controlling 
the LO strength of the tidal deformation of body $A$ by its companion $B$, $R_A$ is 
the star radius and $h_2^{A}$ is the shape Love number~\cite{Damour:2009vw,DL10}.
The dimensionless quantity $h_2^A$ is of order unity, but one is expecting it to be
a function of the relative separation $r$, that increases as $r$ decreases (this was
found in a related black-hole study~\cite{DL10}). In Ref.~\cite{Baiotti:2011am} it 
was found that the effective value $h_2^{\rm eff}=3$ was necessary to allow this 
EOB-predicted contact to occurr always before the NR-defined merger. 
We shall briefly comment  in Sec.~\ref{sec:res} about the magnitude 
of the amplification needed on $h_2$ so to reconcile the EOB contact defined 
by Eq.~\eqref{eq:eob_contact} with a certain NR-defined contact.

\ifusesec
\section{Numerical simulations}
\else
\paragraph*{Numerical simulations}
\fi
\label{sec:nr}

\begin{table}[t]
  \caption{ \label{tab:grid} 
    Summary of the grid configurations and of the runs.
    Columns: %%
    name of the configuration, %%
    maximum refinement level, %%
    minimum moving level, %%
    number of points per direction in the moving levels, %%
    resolution per direction in the level $l=l_{\rm max}$, %%
    number of points per direction in the non-moving levels, %%
    resolution per direction in the level $l=0$. }
  \centering  
  \begin{ruledtabular}  
  \begin{tabular}{ccccccccccc}        
    \hline
    run & $l_{\rm max}$ & $l^{\rm mv}$ & $N_{\rm xyz}^{\rm mv}$ &
    $h_{\rm l_{\rm max}}$ & $N_{\rm xyz}$ & $h_0$ \\
   \hline
    L & 7 & 4 & 100 & 0.1875 & 160 & 24    \\
    M & 7 & 4 & 128 & 0.1466 & 176 & 18.75 \\
    H & 7 & 4 & 160 & 0.1172 & 212 & 15    \\
    \hline
  \end{tabular}
  \end{ruledtabular}  
\end{table}

\begin{figure}[t]
  \begin{center}
    \includegraphics[width=0.49\textwidth]{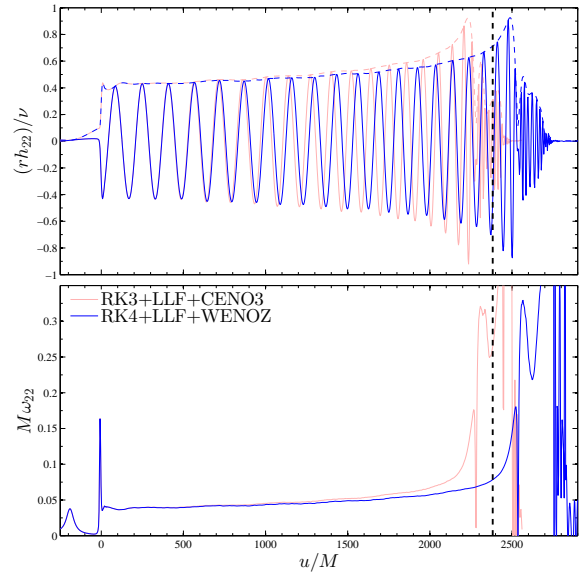}
    \caption{\label{fig:wave} (color online) 
      Numerical quadrupolar gravitational waveform extracted at the outermost radius $r_{\rm obs}=750=247.55M$
      for the CENO and WENO data (run H). Top: real part and amplitude (dashed lines). Bottom: frequency.
      The vertical lines mark the contact of the WENO data.}
  \end{center}
\end{figure}

\begin{figure}[t]
  \begin{center}
    \includegraphics[width=0.49\textwidth]{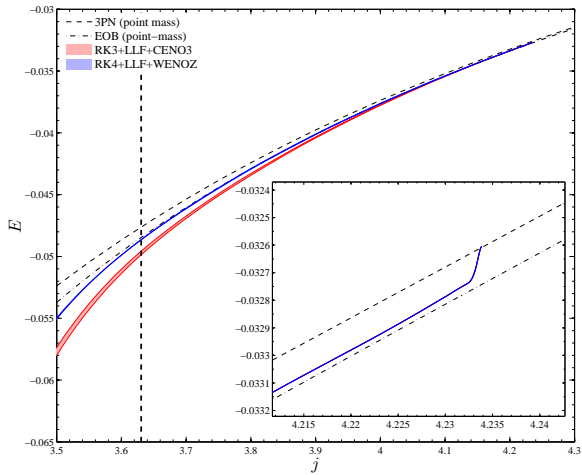}
    \caption{\label{fig:NR_Ej} (color online) 
      Numerical dynamics: $E(j)$ curves for two series of simulation 
      (CENO and WENO data), and for two point-mass analytical models 
      (EOB resummed and ``canonical'' Taylor-expanded 3PN).
      The vertical dashed line marks the angular momentum at
      contact.}
  \end{center}
\end{figure}

\begin{figure*}[t]
  \begin{center}

    \includegraphics[width=0.3\textwidth]{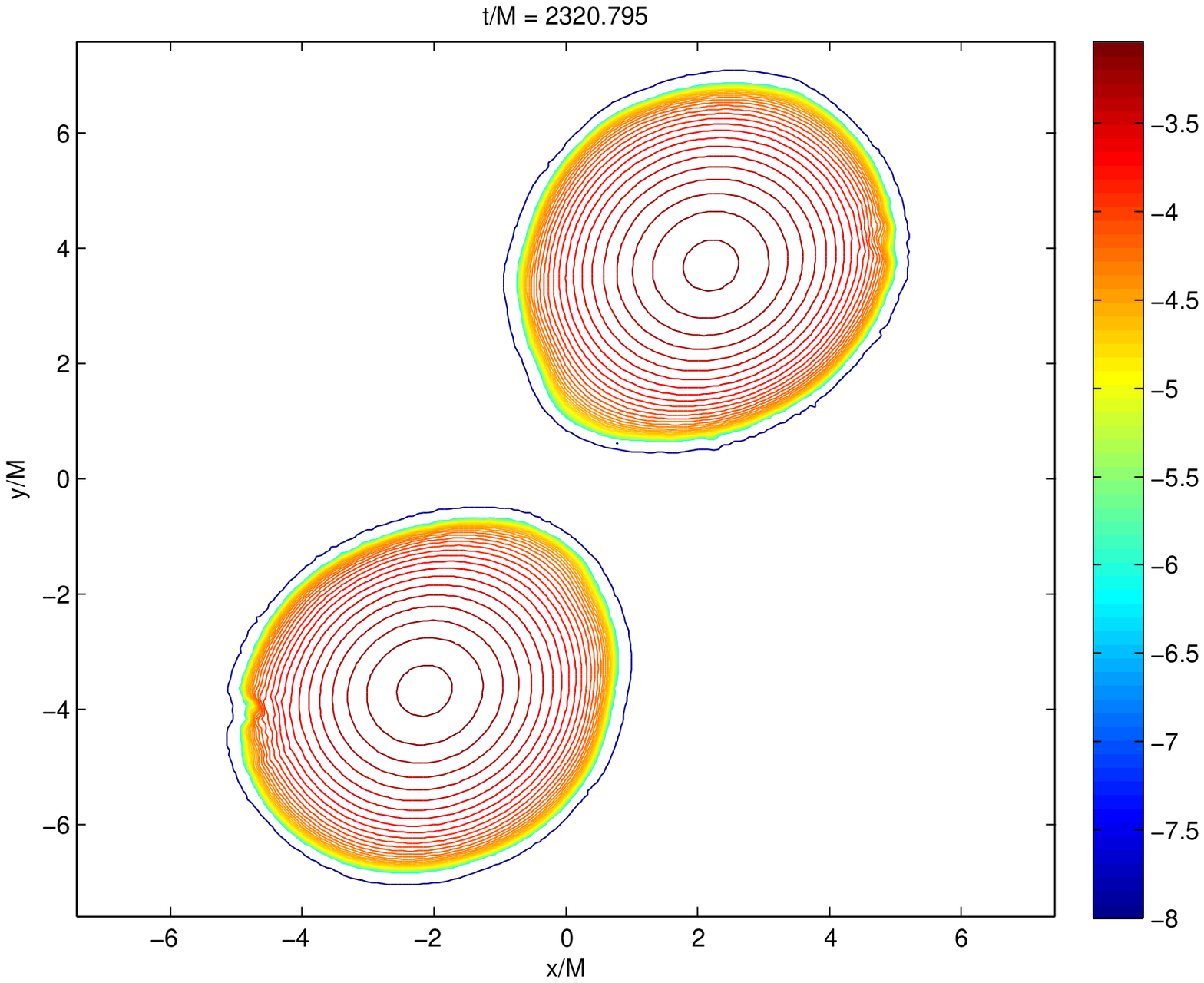}
    \includegraphics[width=0.3\textwidth]{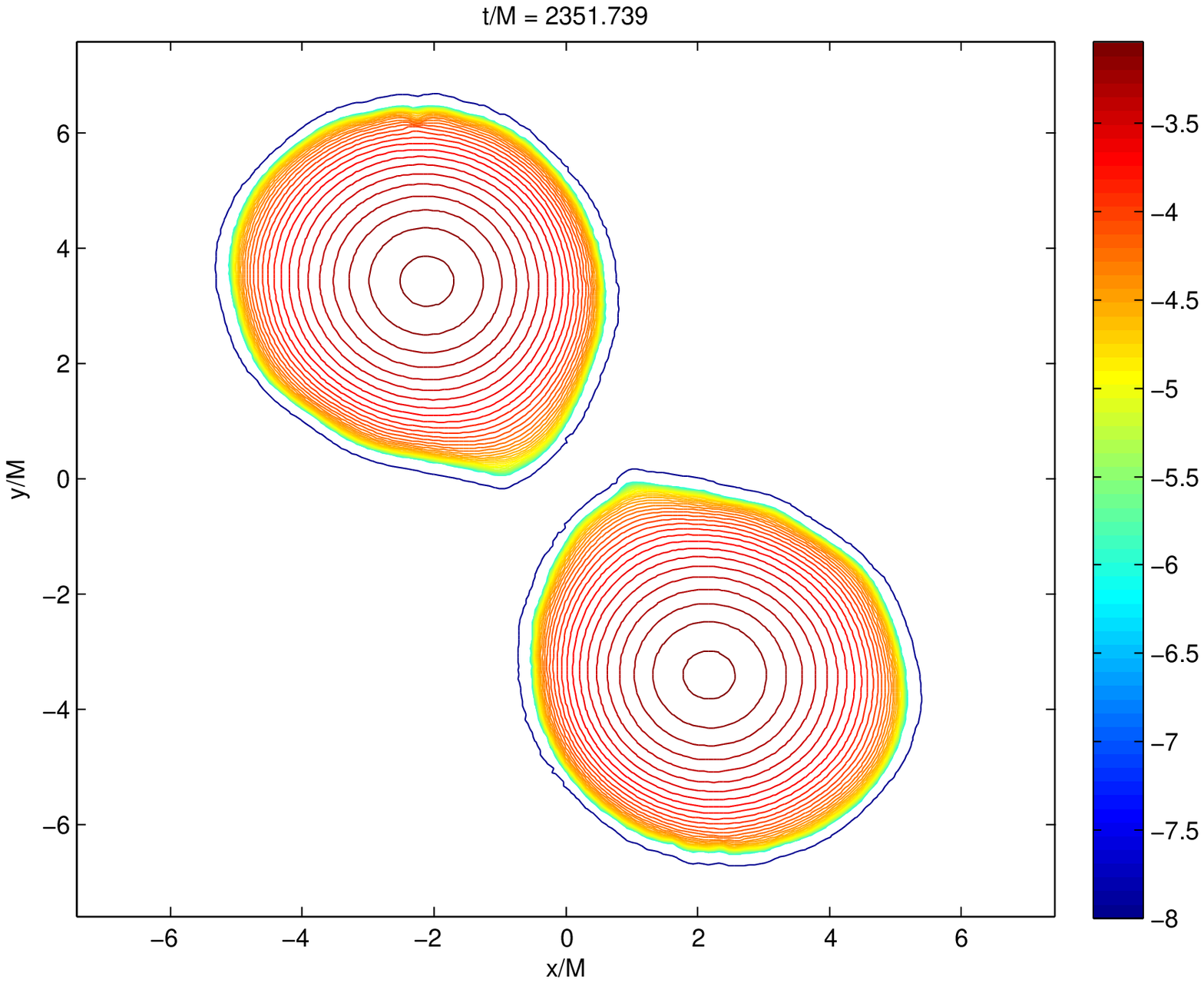}
    \includegraphics[width=0.3\textwidth]{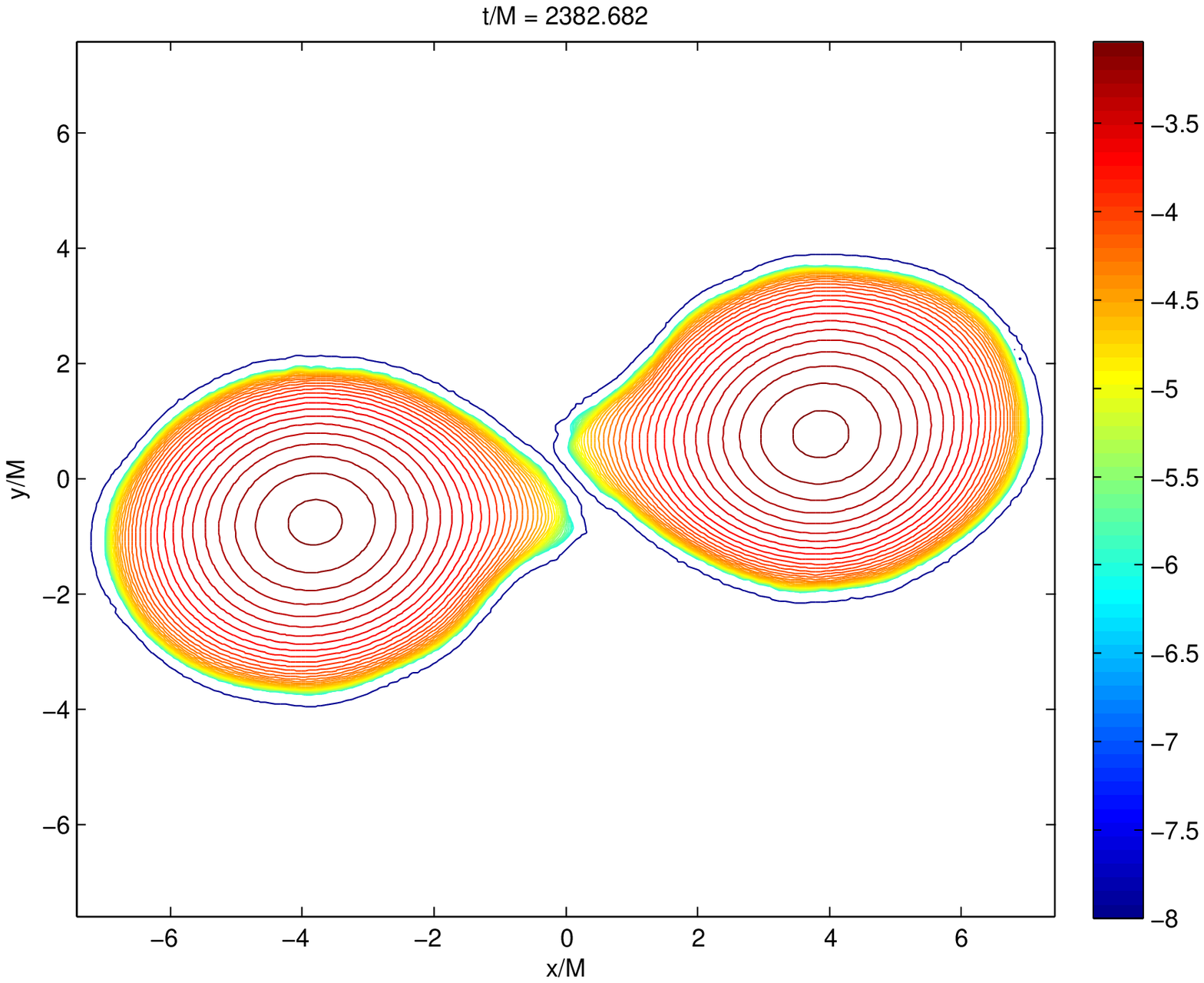}\\
    \includegraphics[width=0.3\textwidth]{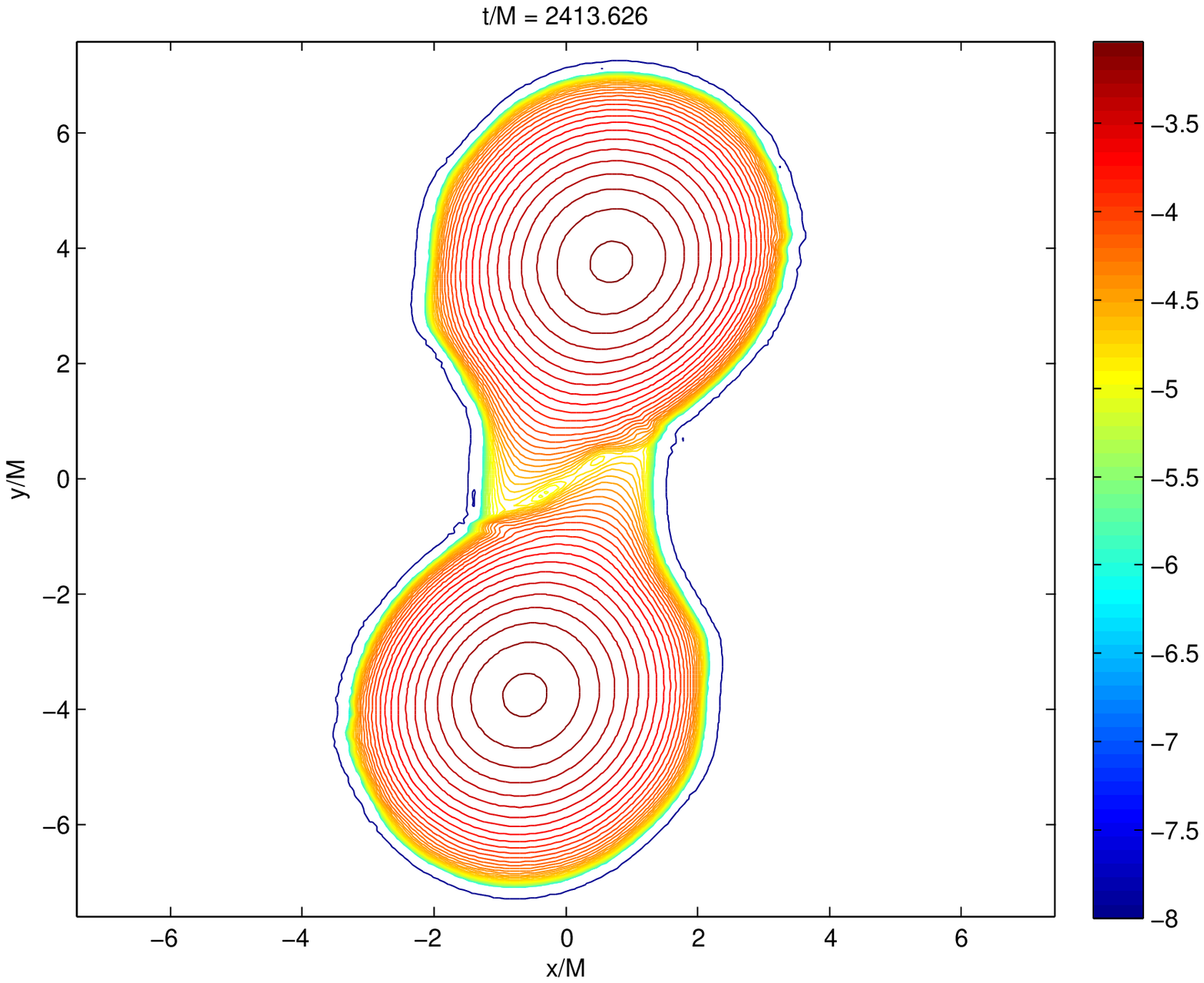}
    \includegraphics[width=0.3\textwidth]{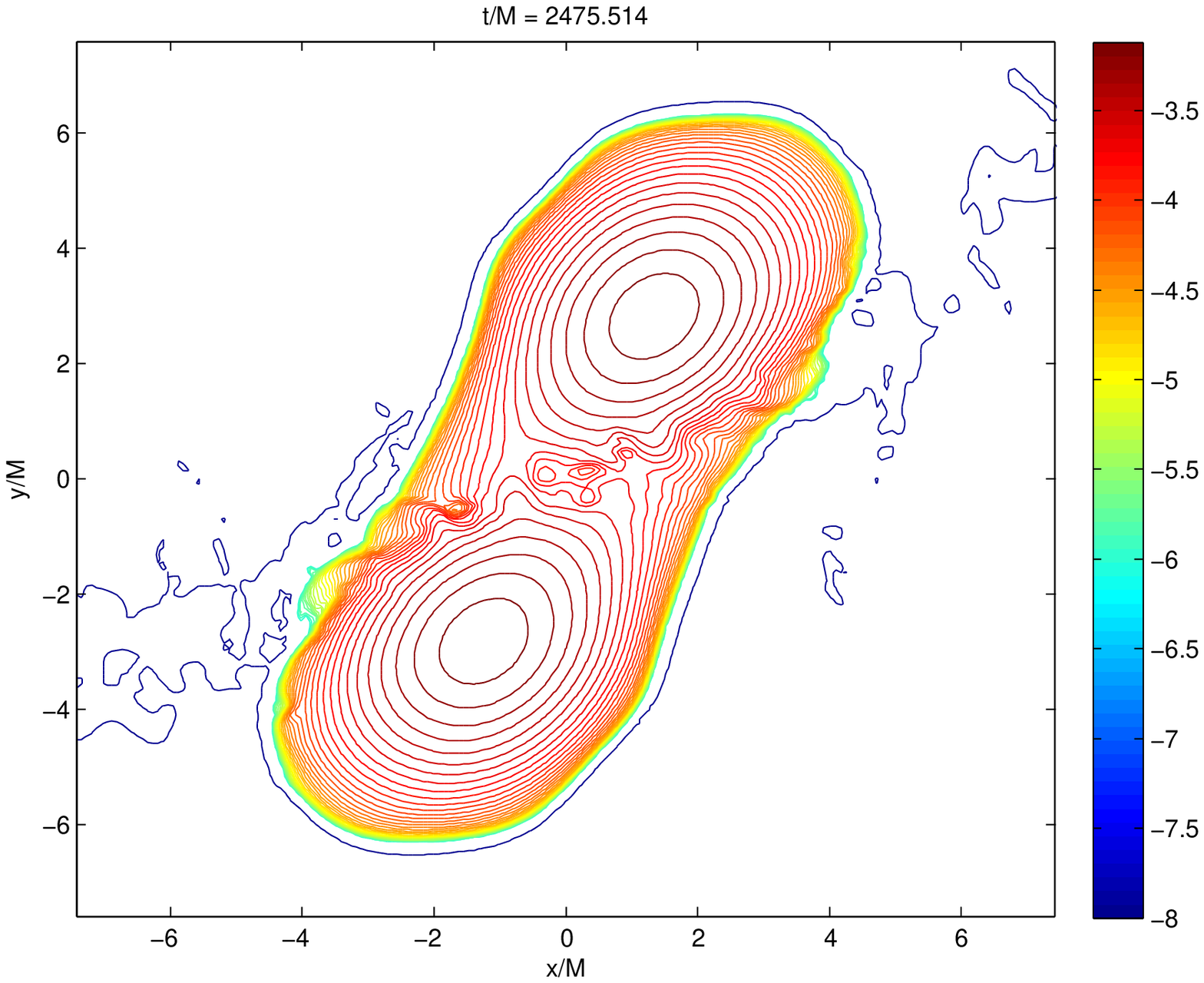}
    \includegraphics[width=0.3\textwidth]{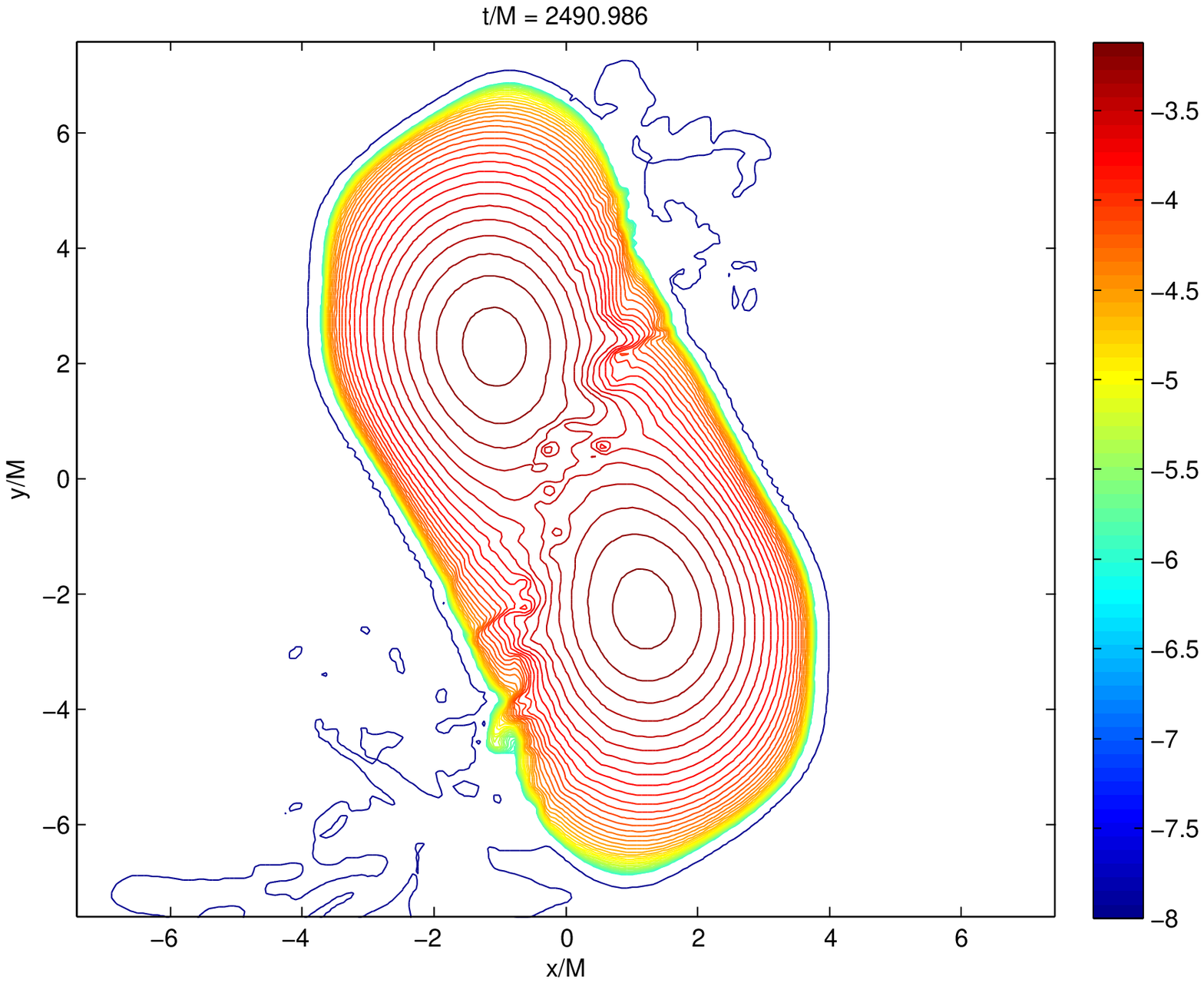}

    \caption{\label{fig:NR_dyn} (color online) Binary dynamics from
      WENO NR simulations. Contour plot of the rest-mass density on
      the equatorial plane. The snapshots indicate the contact happens
      around dynamical time $t_{\rm c}\approx 2382M$. This dynamical
      time, corresponding to observer's retarded time $u_c=t-r_*^{\rm obs}$,
      locates the contact at GW frequency $M\omega^{\rm
        c}_{22}\approx 0.078$. Run H.}  
  \end{center}
\end{figure*}

Target waveforms for the comparison with different analytical
predictions are computed via NR simulations like those presented in
great detail in~\cite{Bernuzzi:2011aq}.  While NR simulations of BNS
have reached a certain maturity (see~\cite{Faber:2012rw} for the most
recent review), intrinsic difficulties in the treatment of general
relativistic hydrodynamics (GRHD) make the numerical calculations of
small effects in long simulations still very challenging.  Note that
the study of tidal effects in the late inspiral requires to resolve
(at least) dephasing of $\lesssim0.5~rad$ over ten cycles.
In particular the numerical viscosity of
high-resolution-shock-capturing scheme (HRSC) typically employed, is
strongly dependent on the reconstruction scheme (cell-interfaces
interpolation), and plays an important role in the accuracy of the
simulations, see e.g.~\cite{Giacomazzo:2010bx,Thierfelder:2011yi}.
While partially under control in short (three orbits) runs by the use
of 3rd order reconstructions, long term simulations such those
presented here are challenging, also due to the computational cost of
extensive testing.

In this work we consider long-term evolutions ($\sim$10 orbits) 
of an equal-mass conformally-flat (CF) and irrotational initial 
configuration of Arnowitt-Deser-Misner (ADM) mass $M_{\rm ADM}^0=3.00506(2)$ and
angular momentum $J_{\rm ADM}^0=9.716(1)$.  The initial separation is
$d\sim50$ associated with GW frequency of $\sim 394~Hz$. The fluid is
described by a $\Gamma$-law EOS ($\Gamma=2$) and isentropic evolution
were considered as in~\cite{Bernuzzi:2011aq}.  The baryonic mass of
each star is $M_{\rm b}=1.62500(0)$, the gravitational mass of each
star in isolation is $M/2=M_A=M_B=1.51483(7)$, radius and compactness
are respectively $R=10.82065(0)$ and ${\cal C}=0.14$.  The
corresponding $\ell=2$ dimensionless Love number is $k_2=0.07890(1)$
and the $\ell=2$ shape Love number is $h_2=0.8699$. 
The initial configuration~\cite{Taniguchi:2002ns} is computed with the
LORENE library and publicly available, and was already considered in
Refs.~\cite{Baiotti:2010xh,Baiotti:2011am,Bernuzzi:2011aq}.

Evolutions were performed with the BAM code described
in~\cite{Bruegmann:2003aw,Brugmann:2008zz,Thierfelder:2011yi}. Here we
mention the GRHD is handled with 
finite-differencing HRSC based on primitive reconstruction, the
Local-Lax-Friedrichs (LLF) central scheme for the numerical fluxes and
Runge-Kutta time integrators, see e.g.~\cite{Zanna:2002qr}. Cartesian
grids and Berger-Oliger adaptive-mesh-refinement (``moving boxes''
technique) are used. The grid setup, resolutions, gauge parameters,
and finite differencing stencils for the metric sector are exactly the
same as the convergent series discussed in~\cite{Bernuzzi:2011aq},
they are listed for completeness in Table~\ref{tab:grid}.

We point out here that focusing on isentropic evolutions is justified
by the following facts: 
(i)~physically, BNS evolutions are expected to be isentropic up to contact;
(ii)~any analytic model can not describe non-isentropic effects
(e.g.~shock heating); 
(iii)~previous works demonstrated~\cite{Thierfelder:2011yi,Baiotti:2011am} that
considering non-isentropic evolutions actually leads to {\it smaller} tidal effects.

Two series of simulations were performed: one is the convergent
series presented in~\cite{Bernuzzi:2011aq}, where the HRSC employs the
(formally) 3rd order convex-essentially-non-oscillatory (CENO3) for
primitive reconstruction and a 3rd order Runge-Kutta scheme. The
second series is computed with the same setup except the use of the
(formally) 5th order weighted-essentially-non-oscillatory (WENOZ)
method of~\cite{Borges20083191} and a Runge-Kutta scheme of 4th order.
In~\cite{Bernuzzi:2011aq} it was presented a detailed analysis of the
uncertainties that affect the waveform due to truncation and
finite-extraction errors; the new data computed for this work show
analogous features. However, we observed differences between the two
data sets. For a given resolution, the merger in the CENO
data occurs earlier than in the WENO data. The dominant multipole
$\ell=m=2$ of the metric waveforms, $h_{22}$, obtained from 
the two different setup is displayed in Fig.~\ref{fig:wave}. In
abscissa we use the retarded time $u\equiv t -r_*$, where $r_*=r_{\rm
  S} + 2M_{\rm ADM} \log(r_{\rm S}/(2M_{\rm ADM})-1)$, and $r_{\rm S}$
is the Schwarzschild radius corresponding to the isotropic
(coordinate) radius $r$.  The waveforms are extracted at the 
outermost radius $r=750=247.55M$. The simulations compute waveforms
from the Newman-Penrose scalar $\psi_4$, that
is then decomposed in spherical harmonics modes, $\psi_4^\lm$.
The metric multipoles $h_\lm$ are calculated from the $\psi_4^\lm$ 
by integrating the relation $\psi_4^\lm=\ddot{h}_\lm$. To do the  
integration, we use the frequency-domain procedure of~\cite{Reisswig:2010di} 
with a low-frequency cut off at $\omega_0=0.02/M$. 
The signal is integrated from the very beginning of the simulation, in 
order to include also the initial burst of radiation related  to the
use of CF initial data. This radiation is often called (somehow
improperly) ``junk'' radiation. 
Note that in the text for brevity we consider the metric waveform
multiplied by the extraction radius without explicitly changing the
notation, i.e.~$h_\lm\equiv r h_\lm$.
As it is clear from the figure, at a given resolution, CENO and WENO
waveforms accumulate a significant relative dephasing towards merger;
uncertainties due to the HRSC numerical viscosity become larger as the 
simulation time advances and eventually dominant over truncation 
(and finite extraction) errors towards contact ($M\omega_{22}\sim0.07$~\cite{Bernuzzi:2011aq}, 
see below for an estimate of the GW frequency of the contact), where any 
convergent behavior is lost. For both data sets the higher the resolution, 
the later is the merger~\cite{Thierfelder:2011yi,Bernuzzi:2011aq}.  
In practical terms, Fig.~\ref{fig:wave} indicates that the 
differences in the HRSC effectively influence the magnitude of the
tidal interaction between the two stars, from a larger value for the
CENO data to a smaller one for the WENO data.

This effect can be properly quantified by exploring the actual
dynamics of the BNS system so to contrast it with the corresponding
point-mass one. An intrinsic, gauge-invariant, way of doing so is
by means of the relation between the total energy $\E$ and total angular
momentum $\J$ of the system, $\E(\J)$. Following Ref.~\cite{Damour:2011fu} that 
computed this quantity for binary black hole (BBH) systems, $\E$ and $\J$ 
are obtained from the NR data as 
\be
\label{eq:E} \E^{\rm NR} (u) = \E_0 - \Delta \E_{\rm rad}^{\rm NR} (u) \, , \ee
\be
\label{eq:J} \J^{\rm NR} (u) = \vert \mathbfcal{J}_0 - \Delta
\mathbfcal{J}_{\rm rad}^{\rm NR} (u) \vert \, . \ee 
Here,
$\E_0=M^0_{\rm ADM}/M$ and $\J=J^0_{\rm ADM}/M^2$ are the initial ADM
mass and angular momentum expressed in units of the total
gravitational mass of the stars in isolation; $\Delta \E_{\rm rad}^{\rm NR}$
and $\Delta \J_{\rm rad}^{\rm NR}$ (expressed in the same units) are the
radiated energy and angular momentum between the initial (retarded)
time $u_0$ and $u$. They are computed from the multipole moments of
the metric waveform $h_\lm$ and of its time derivative $\dot{h}_\lm$,
as
\begin{align}
\Delta \E_{\rm rad}^{\rm NR}(u) &=\dfrac{1}{16\pi}\sum_{\ell=2}^{\ell_{\rm max}}
\sum_{m=0}^{\ell}\int_{u_0}^u du'\left|\dot{h}_\lm(u')\right|^2\\
\Delta \J_{z\rm rad}^{\rm NR}(u)&=\dfrac{1}{16\pi}\sum_{\ell=2}^{\ell_{\rm max}}\sum_{m=1}^{\ell}
\int_{u_0}^{u}d u' m
\Im\left[h_\lm(u')\dot{h}_\lm^*(u')\right].
\end{align}
In Eq.~\eqref{eq:J} we also included the
$x$, and $y$ component of the angular momentum loss, $\Delta\J^{\rm
NR}_x$ and $\Delta\J^{\rm NR}_x$, though, as expected, they are of
order $10^{-10}$ and negligible in practice. For convenience, we work
with the binding energy per reduced mass $E \equiv (\E-M)/\mu$ and the
dimensionless rescaled angular momentum $j\equiv \J/M\mu$, where
$\mu=M_A M_B/M$.

The main panel of Fig.~\ref{fig:NR_Ej} compares the NR relations
$E^{\rm NR}(j)$ computed for the two data series (CENO and WENO)
with two analytical, point-mass, curves: the canonical PN expanded 
$E(j)$ relation (see Eq.~(5) of~\cite{Damour:2011fu}), dashed line, 
and the NR-tuned, EOB resummed one (dash-dotted line), that was 
found in~\cite{Damour:2011fu} to show an excellent agreement with 
corresponding BBH numerical curve~\footnote{ %%
  Although the canonical PN curve was very close to the CF 
  initial state, it was found to progressively deviate from the
  numerical calculation, giving then an inaccurate representation of
  the point-mass dynamics.}  
(see Fig.~2 of~\cite{Damour:2011fu}, top panel). 
The $E^{\rm NR}(j)$ curves used here were obtained
from waveforms taken at the the outermost extraction radius,
$r_{\rm obs}=750=247.55M$, and, for simplicity, by including  only 
the $\ell=m=2$ multipole in the calculation of Eqs.~\eqref{eq:E}-\eqref{eq:J}. 
To illustrate the influence of the uncertainties due to finite resolution, 
instead of displaying the NR data as a simple curve, 
we present them as the shaded band that is included between the medium 
(run M, bottom border) and high (run H, top border) resolutions.  
The diagram illustrates that, while the
CENO data graze the EOB point-mass curve at the very beginning of the
simulation (see inset) they visibly deviate from it after, indicating
the presence of strong tidal effects. On the contrary, the WENO data
remain always very close to the point-mass EOB curve, so to be almost
indistinguishable on the scale of the main plot.

We experimentally conclude that the more dissipative numerical setup
(CENO data) artificially amplifies tidal effects, leading to severe
inaccuracies on the waveform phasing.
The improvements obtained in the WENO data and the availability of
convergence tests and error estimates~\cite{Bernuzzi:2011aq} are
crucial for the comparison with the analytical information presented
below. 

Let us finally briefly comment the inset of Fig.~\ref{fig:NR_Ej},
which focuses only on the initial part of the $E^{\rm NR}(j)$ curve. Similarly
to the black-hole case (see inset of Fig.~1 of~\cite{Damour:2011fu}),
the initial state of the system is very close to the point-mass, 3PN
canonical $E(j)$ curve.  Then the effect of the losses due to the
junk radiation moves the initial state down, close to the EOB
curve. Note in addition that this early-inspiral part the 
NR curve is {\it above} the point-mass EOB curve because only the $\ell=m=2$ 
mode has been included. We shall explore the effect of the other multipoles 
in the next Section, in the context of the detailed comparison 
with the tidal EOB model.

For the comparison with analytical predictions, it is important to have
an estimate of some {\it contact} frequency extracted from the
NR data, since the tidal EOB model ceases to be valid after. 
However, connecting the (local, strong-field) dynamics, parametrized by the dynamical
time $t$, of the two objects with the radiation observed in the wave-zone unambigously,
is a non-trivial task. In first approximation, any phenomenon occurring in the strong 
field region at dynamical time $t$ reaches the observer $r_{\rm obs}$ at retarded 
time $u=t-r_*(r_{\rm obs})$. For simplicity, in the following we shall 
assume such relation to connect the two events, though, in doing so, 
we are neglecting an additional time-delay due to the propagation of the 
signal in the strong-field region.
In Fig.~\ref{fig:NR_dyn} we show snapshots of the rest-mass density in
the orbital plane at few dynamical times around the contact of the two
stars. The pictures indicates clearly that at $t_{\rm c}\approx 2382M$ 
the well-known shearing contact is taking place. The
corresponding GW frequency is $M\omega_{22}(u_c=2382M)\approx 0.078$, and the
corresponding value of the angular momentum is $j_{\rm c}\approx 3.63$. 
The latter is shown in the dashed vertical line in Fig.~\ref{fig:NR_Ej}. 
Note that the final merger (e.g, formally defined by the first peak of 
$|h_{22}|$) occurs later, at $M\omega_{22}\sim 0.13$.

\ifusesec
\section{EOB/NR comparison}
\else
\paragraph*{EOB/NR comparison}
\fi
\label{sec:res}

In this Section we consider only the WENO data and compare them with
the tidal EOB model. We present two types of comparisons, one for the
dynamics, through the $E(j)$ curve, and one for the phasing.
We shall take as ``best'' (multipolar) waveform the one computed with the
highest resolution available and extracted at the outermost radius
$r_{\rm obs}= 247.55M$. 

\ifusesec
\subsection{Dynamics}
\else
\fi

\begin{figure*}[t]
  \begin{center}
    \includegraphics[width=0.45\textwidth]{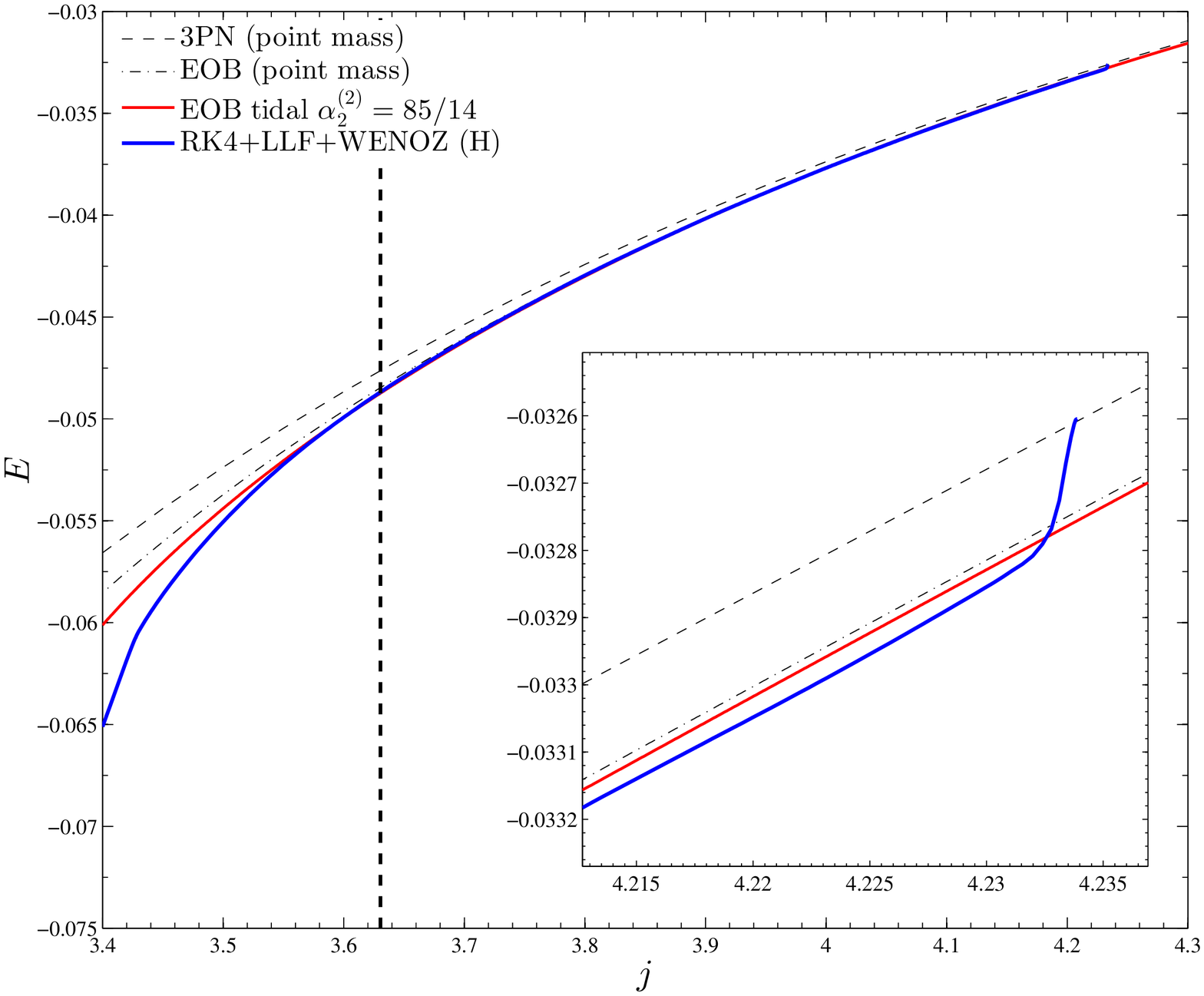}
    \includegraphics[width=0.45\textwidth]{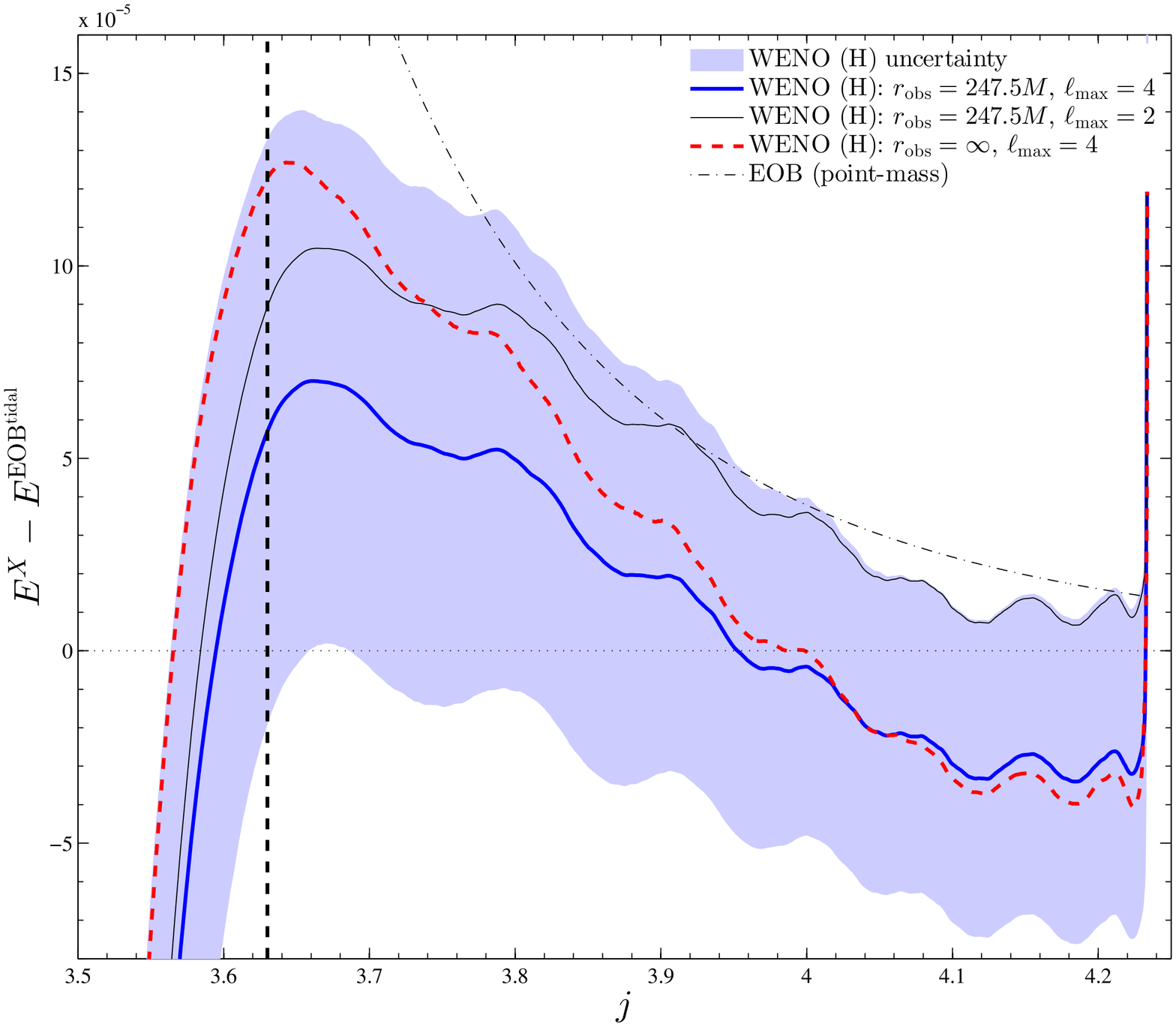}
    \caption{\label{fig:E_Vs_j} Comparison between EOB and NR dynamics. Left panel: 
    reduced binding energy ($E$) versus reduced angular momentum ($j$) curves.
    Right panel: differences with the tidal EOB model. The shaded region represents
    the estimated uncertainties on the ``best'' numerical curve. Numerical data are consistent
    with the tidal EOB model.}
  \end{center}
\end{figure*}

Let us first compare the NR $E^{\rm NR}(j)$ relation to the
corresponding analytical prediction.  This is done in the left panel
of Fig.~\ref{fig:E_Vs_j}. Together with the point-mass analytical
curves (EOB and canonical 3PN) already shown in Fig.~\ref{fig:NR_dyn}, 
we also display the analytical $E(j)$ EOB curve that includes tidal 
effects (solid, red online) at NNLO order. The numerical curve 
(solid, black online) is different from the corresponding band of 
Fig.~\ref{fig:NR_dyn} in that $\Delta \J^{\rm
NR}_{\rm rad}$ and $\Delta \E^{\rm NR}_{\rm rad}$ include all
multipolar contributions up to $\ell_{\rm max}=4$.  Note that,
although the symmetry of the system implies that all multipoles with
odd values of $m$ have to be zero, they are actually nonzero, and are
included in the computation of the losses to which they contribute as
very small amplitude, structureless, noise. In these simulations,
differently from~\cite{Baiotti:2011am}, we do not impose rotational
(``$\pi$'') symmetry on the orbital plane, but evolve instead the
equations in the whole plane $z>0$.  To give numbers in the $\ell=2$
case, during the inspiral it is, at maximum, $|\dot{h}_{22}|\sim
5\times 10^{-3}$, $|\dot{h}_{20}|\sim 2\times 10^{-5}$ and
$|\dot{h}_{21}|\sim 6\times 10^{-6}$.

Though the $m=0$ modes are practically negligible during the inspiral,
they actually contribute to the initial burst of radiation and thus
they must be included in the computation of $E^{\rm NR}(j)$. For
example, during this epoch, that lasts for $\sim 25M$, 
$|\dot{h}_{20}|\sim 0.5\times |\dot{h}_{22}|$ and
$|\dot{h}_{40}|\sim 0.5 \times |\dot{h}_{44}|$.
By contrasting the insets of Fig.~\ref{fig:NR_dyn} and of the left-panel 
of Fig.~\ref{fig:E_Vs_j}, we observe that the subdominat multipoles drive 
the NR curve not only below the point-mass curve, but even below the 
tidal EOB curve~\footnote{ 
  The tidal EOB curve is below the corresponding
  point-mass one indicating that, due to the attractive nature of tidal
  interaction, the system is gravitationally more bound.}.
Like in Fig.~\ref{fig:NR_dyn}, the dashed vertical line marks 
the location of the NR-contact.
We note in passing that to make the EOB-contact consistent with 
the NR-defined contact we should replace in Eq.~\eqref{eq:eob_contact} the 
value $h_2=0.8699$ (which would give a EOB contact frequency slightly smaller 
than the merger frequency~\footnote{Similarly, the ``bare'' contact, with $h_2=0$,
is even closer to the merger, giving $M\omega_{22}\sim 0.106$.} 
$M\omega_{22}\sim 0.094$) with $h_2^{\rm eff}\sim 3.4$, 
which yields an EOB contact frequency $\sim 0.078$. This estimate of the 
deformation of each star coming from actual data is consistent with the guess 
of Ref.~\cite{Baiotti:2011am} ($h_2^{\rm eff}=3$), but further
work will be needed to understand, numerically, the actual
amplification experienced by $h_2$.  

By close inspection of the plot we observe that the NR curve actually
change its slope: around $j\sim3.8$ it lies between the tidal and
point-mass curves, while during contact the curve bends again below
the tidal EOB one.  We argue this behavior is caused by numerical
inaccuracies, for example small violations of mass conservation, to
which the computation of $E^{\rm NR}(j)$ is extremely sensitive. 
It was not possible to identify the precise cause, but we mention that it
corresponds to a numerical oscillation at the fourth digit of $E^{\rm NR}(j)$. 

The differences between numerical data and the analytical predictions
are made precise in the right panel of Fig.~\ref{fig:E_Vs_j}. The four
curves represent the four differences $E^X-E^{\rm EOB_{\rm tidal}}$
where the label $X$ indicates in turn: NR (thick, red online) with
$\ell_{\rm max}=4$; NR (thin, black online) with $\ell_{\rm max}=2$;
NR (thick, dashed, black online) with $\ell_{\rm max}=4$ extrapolated
at infinite extraction radius; EOB point-mass (dash-dotted, black
online).  The shaded region represents the estimated uncertainty
affecting the best waveform.  It has been obtained by taking into
account the following three effects: 
(i) resolution; 
(ii) finite extraction radius; 
(iii) contribution of higher multipoles.
To compute it, we first took the following differences 
between $E^{\rm NR}(j)$ curves: 
(i) between H and M data, $\Delta E^{\rm HM}$; 
(ii) between H data extrapolated at infinite extraction radius and
at $r_{\rm obs}=247.55M$, $\Delta E^{\rm H}_\infty$. The extrapolation was done from
data at $r_{\rm obs}=\{400,\, 500,\, 600,\, 700,\, 750\}$, aligned using
retarded time, taking a quadratic fit; 
(iii) between H data with $\ell_{\rm max}=4$ and with 
$\ell_{\rm  max}=2$, $\Delta E_\ell^{\rm H}$.  Then the final (conservative) 
error-bar is obtained as $\Delta^{\rm NR} E(j)=\pm \sqrt{(\Delta E_\infty^H)^2
+ \left(\Delta E^{\rm HM}\right)^2 
+ \left(\Delta E_\ell^H\right)^2}$.  Note that at large values of $j$
(beginning of the simulation) the most relevant uncertainty is the one 
due to the finite radius and to the choice of $\ell_{\rm max}$; after that, 
finite-resolution effects become dominant.

The analysis of the dynamics shows that the NR data are {\it
consistent} with the state-of-the-art tidal EOB model, and
distinguishable from the point-mass EOB model up to contact 
($j_c=3.62$). Remarkably the NR curves stay very close to the 
tidal EOB up to $j\sim3.5$, point at which the differences 
with the point-mass curves are at least two-sigma beyond 
the uncertainties.

\ifusesec
\subsection{Waveforms and phasing}
\else
\fi

\begin{figure*}[t]
  \begin{center}   
    \includegraphics[width=0.45\textwidth]{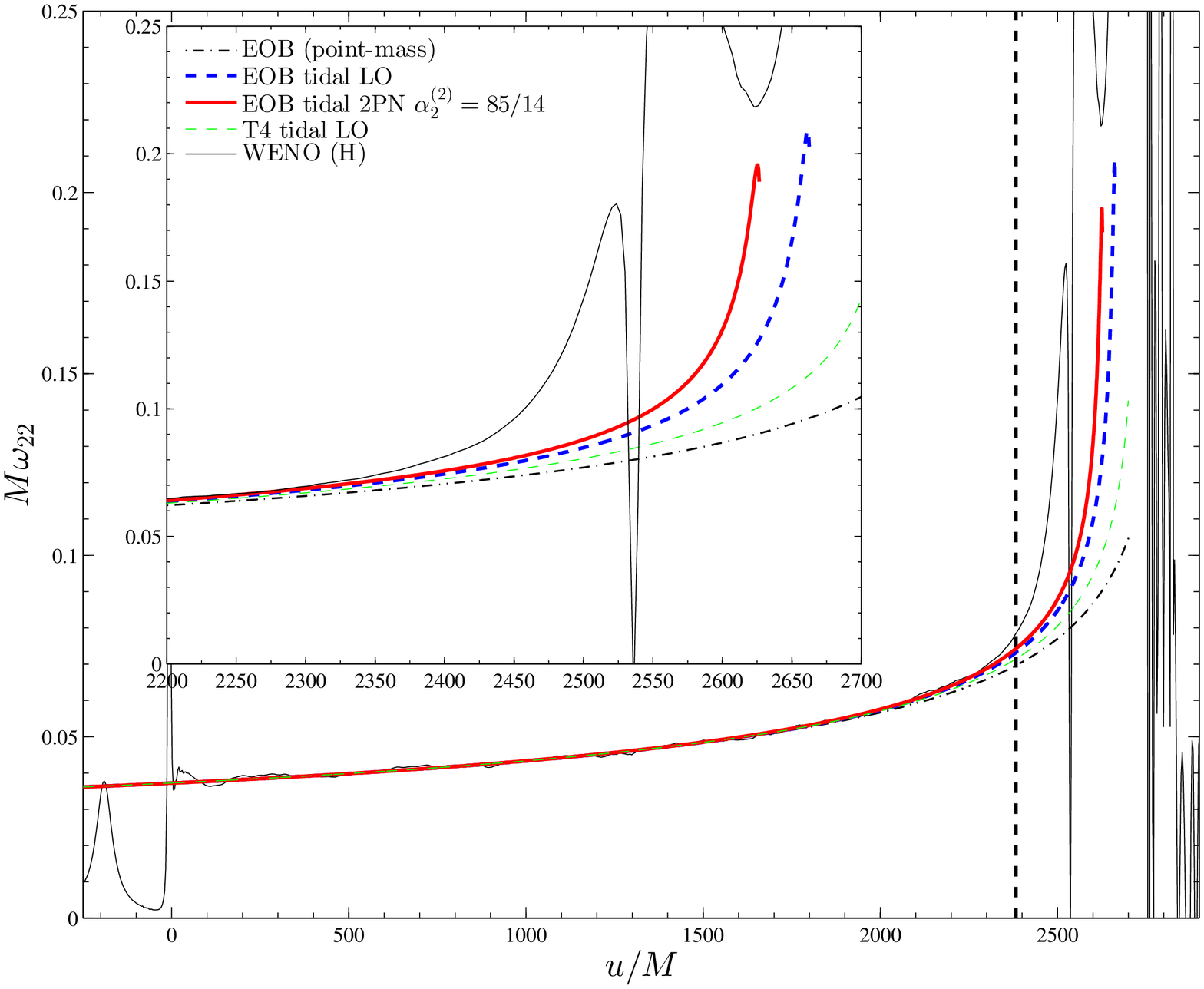}
    \includegraphics[width=0.45\textwidth]{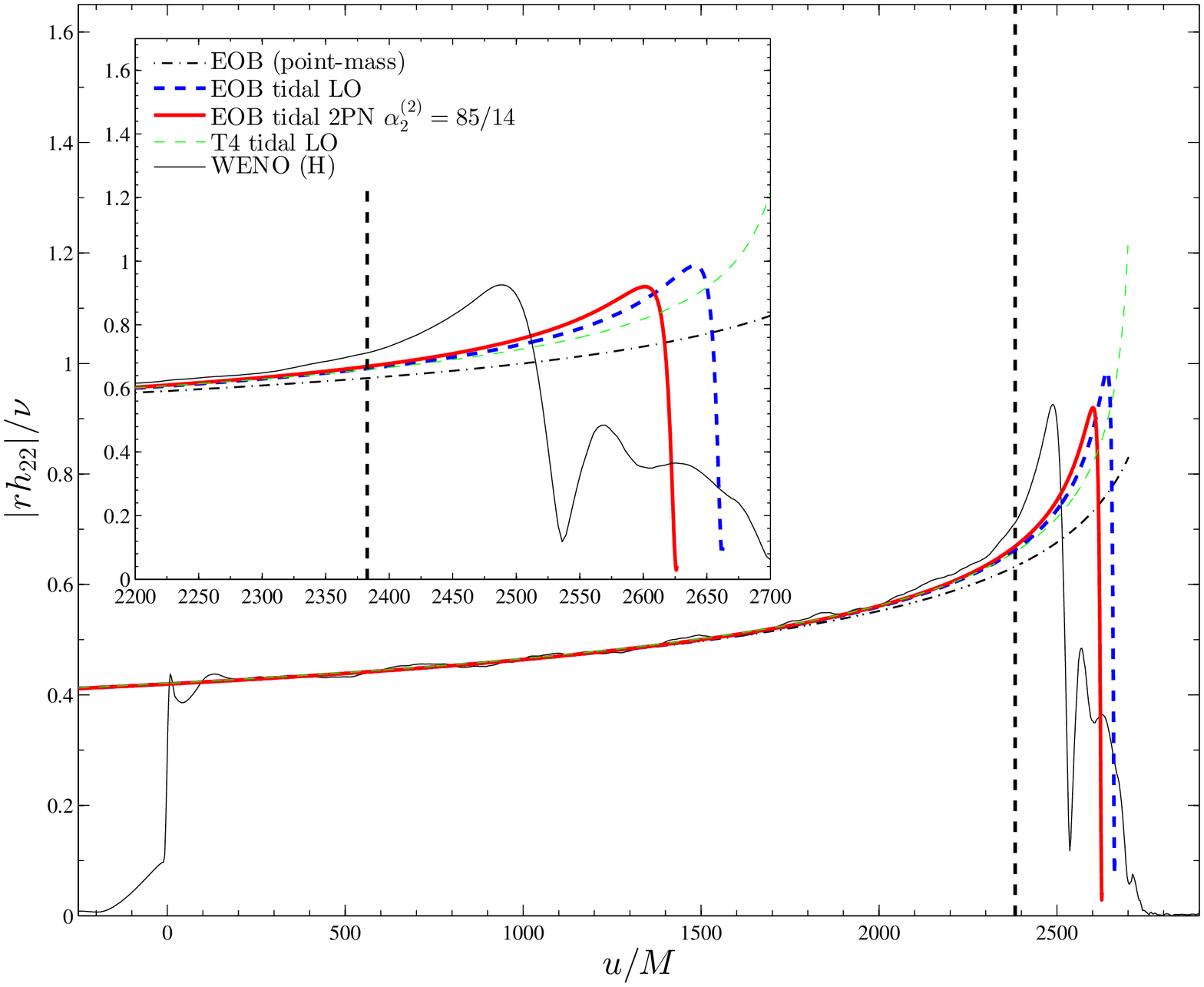}\\
    \includegraphics[width=0.45\textwidth]{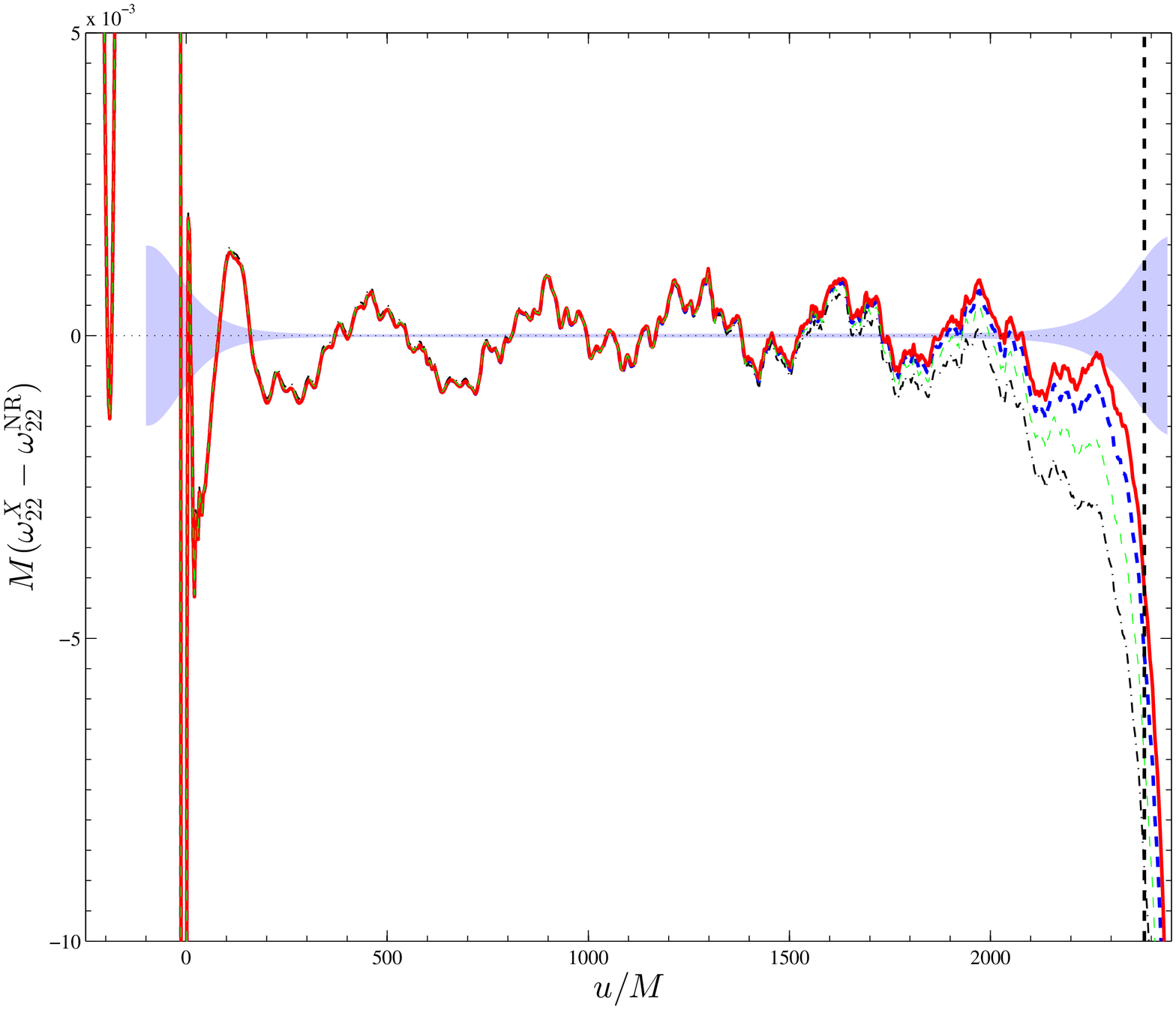}
    \includegraphics[width=0.45\textwidth]{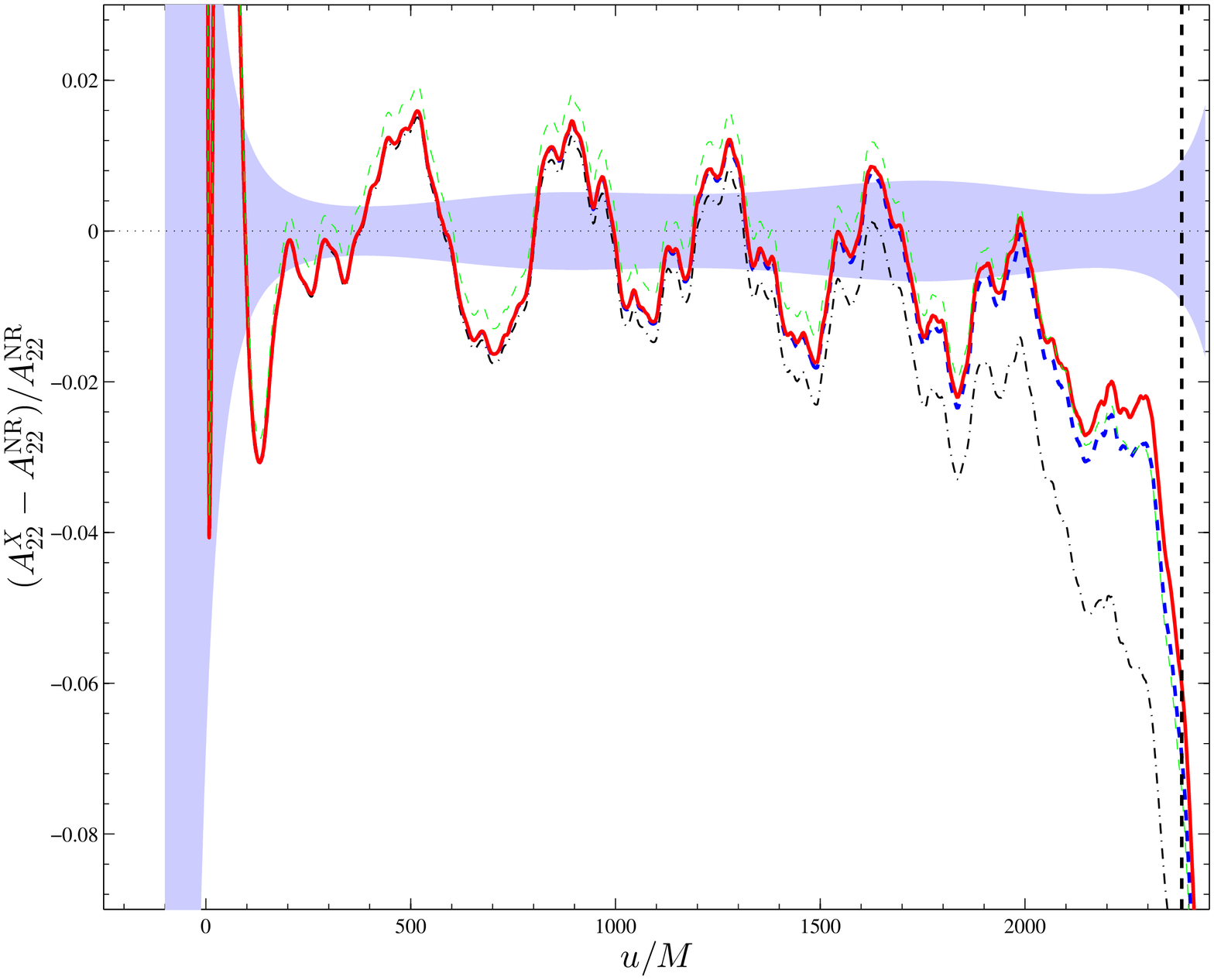}
    \end{center}  
    \caption{\label{fig:freq_mod}
      Comparing numerical and analytical $\ell=m=2$ waveform: 
      frequency (left panels) and modulus (right panels). The vertical line locates the
      NR contact. The shaded regions in the bottom panels are error estimates on the NR
      frequency and modulus.}
\end{figure*}

\begin{figure}[t]
  \begin{center}
    \includegraphics[width=0.5\textwidth]{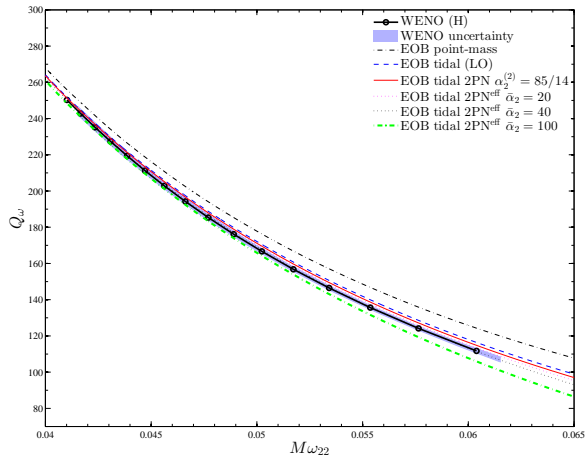}
    \caption{\label{fig:Qomg}(color online). Comparison between various EOB $Q_\omega$ curves and the NR one.
      The good visual agreement between the LO and 2PN tidal EOB models and the NR curve highligths the 
      presence of tidal interaction during the inspiral. Note that the effective value  
      $\bar{\alpha}_2=100$ used in Ref.~\cite{Baiotti:2011am} is incompatible with the NR curve.}  
  \end{center}
\end{figure}

\begin{table}[t]
  \caption{\label{tab:dphi} Phase differences $\Delta\phi =\phi^{\rm EOB}-\phi^{\rm NR}$,
  with its uncertainty $\sigma_{\Delta\phi}$,
  accumulated between frequencies $[\omega_1,\omega_2]=[0.041,\ 0.062]/M$ obtained by
  integrating the difference between the NR $Q_{\omega}$ and some of the EOB curves 
  of Fig.~\ref{fig:Qomg}: the point-mass, the tidal LO and the 2PN
  (NNLO) analytical one with $\alpha_2^{(2)}=85/14$. The error-bar
  $\pm\sigma_{\Delta\phi}$ is obtained by integrating  the shaded
  region around the NR $Q_\omega$ curve in Fig.~\ref{fig:Qomg}.}  
  \centering    
  \begin{ruledtabular}
  \begin{tabular}{cccc}        
    \hline
     EOB model  & point-mass & LO tidal & NNLO tidal \\
     \hline     
     $\Delta\phi$ [$rad$] & 3.92 & 1.49 & 1.06 \\
     $\sigma_{\Delta\phi}$ [$rad$] & 0.61 & 0.61 & 0.61\\
    \hline
  \end{tabular}
  \end{ruledtabular}
\end{table}

\begin{figure}[t]
  \begin{center}
    \includegraphics[width=0.5\textwidth]{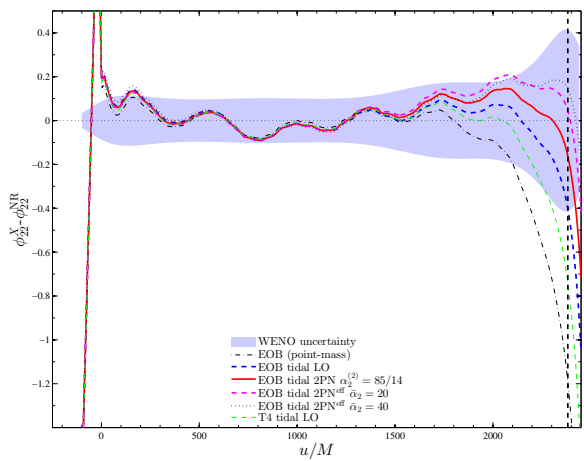}
    \caption{ \label{fig:phasing} (color online) Phase difference between various analytical phase and the NR phase.
      The vertical line indicates the NR contact. The phase difference of $\sim~-0.2rad$ between the 2PN tidal
      EOB model and the NR waveform is compatible with the error estimates on the latter.}  
  \end{center}
\end{figure}

\begin{figure}[t]
  \begin{center}
    \includegraphics[width=0.5\textwidth]{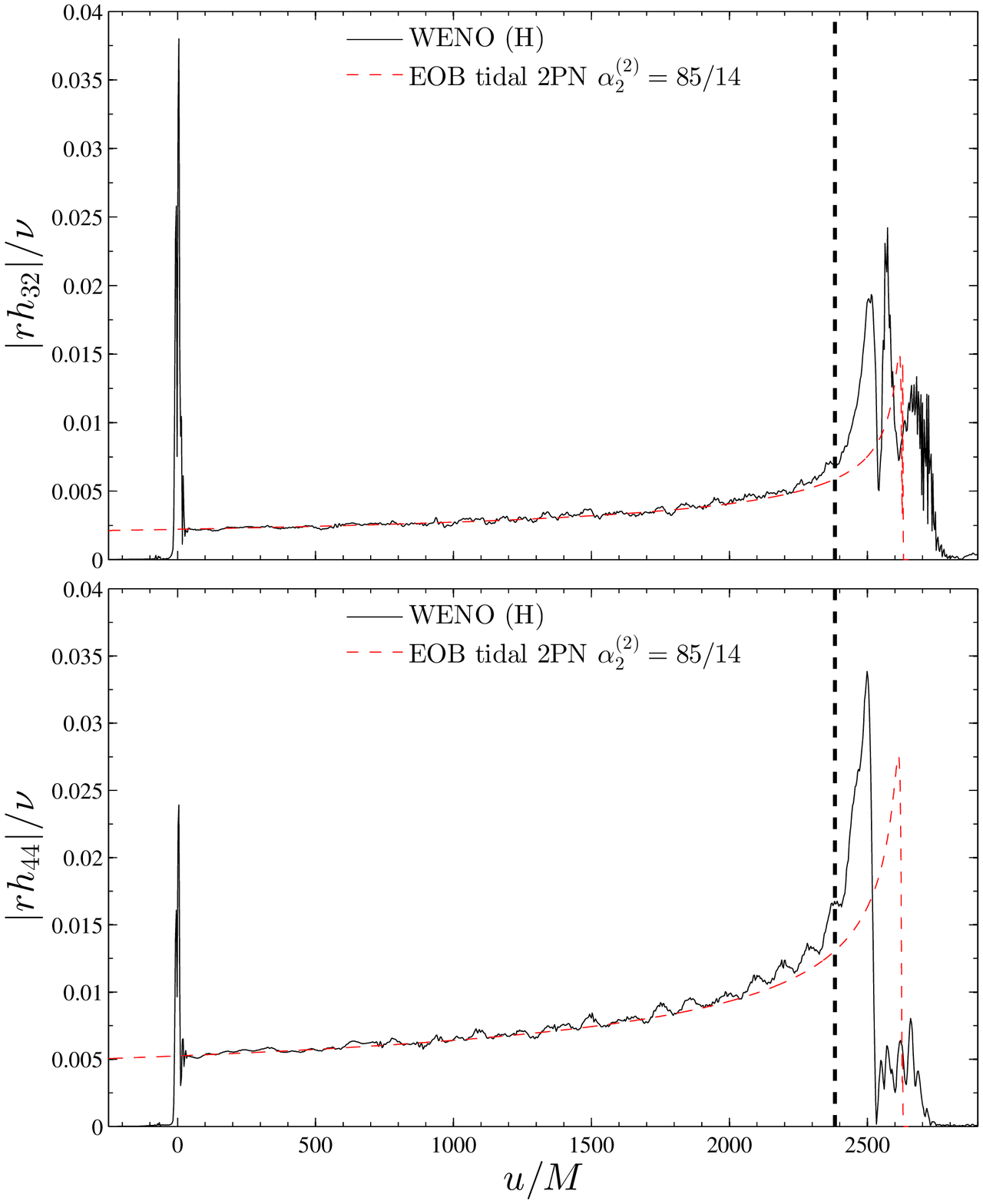}
    \caption{ \label{fig:l3l4} (color online). Comparing EOB and NR waveform moduli for the most relevant
      subdominant multipoles $h_{23}$ and $h_{44}$. The vertical line marks the NR-defined contact time.}
  \end{center}
\end{figure}

Let us now compare the waveforms and quantify the phase difference
accumulated between NR and EOB data in the last orbits of the inspiral.
Focusing on the $\ell=m=2$ mode, we first present a time domain
comparison of the amplitude and frequency, and then switch to a more
quantitative  analysis of the phasing. Higher multipoles are discussed
at the end of the section.

In Fig.~\ref{fig:freq_mod} we contrast the NR waveform
(thin solid line, black online) with 4 different analytical models:
the EOB point-mass (no tides, dash-dotted line), the tidal EOB with
only LO tidal effects (i.e., we set $\hat{A}_\ell^{\rm
tidal}=1$ for all $\ell$'s in Eq.~\eqref{eq:atidal}), dashed, thick
curves, blue online), the 2PN-accurate (NNLO) tidal EOB
model (solid thick curves, red online) and the PN-expanded Taylor T4
model with leading order tidal effects (see Sec.~IIIC
of~\cite{Baiotti:2011am}).  We show together the gravitational
frequency (top left) and waveform modulus (top right) while the
corresponding differences (EOB-NR) are exhibited in the bottom panels.
The vertical dashed line on each panel indicates the NR contact
frequency $M\omega^{\rm c}_{22}\approx 0.075$, after which
we don't expect any analytical model to be accurate.
The relative time, $\tau$ and phase, $\alpha$, shifts that are necessary
to align the analytical to the numerical waveform are determined using
the procedure described in Sec.~VA of Ref.~\cite{Baiotti:2011am}, that
relies on the minimization of the $\chi^2$ of the phase difference
over a certain frequency interval. Here we use the frequency interval
$(\omega_1,\omega_2)=(0.038,0.049)/M$, (corresponding to $(u_1,u_2)/M=(165.8,1610.9)$,
that begins after the initial burst of radiation, so as to remove also
possible inaccuracies due to the integration procedure needed to
get $h_{22}$ from $\psi_4^{22}$. The shaded regions in the bottom panels 
indicate the NR uncertainty, that takes into account finite-extraction
radius and finite resolution effects. It is obtained by: (i) Richardson 
extrapolating  the two highest resolutions assuming second order convergence
and taking the difference with run H data; (ii) similarly, taking the 
difference between the NR waveform extrapolated in extraction 
radius~\footnote{In Ref.~\cite{Bernuzzi:2011aq} it was shown that 
in this case phase difference due to finite-radius effects varies between 
$\sim 0.2$ rad during the early inspiral to $\sim 0.1$ rad at merger, 
while the fractional difference in amplitude varies between $1\%$ and $0.5\%$. } 
and the one at $r_{\rm obs}=750$. In practice, for the first half of the
simulation ($u\sim 1000M$) the uncertainty is dominated by finite-extraction-radius
effects, while later it is the resolution to play the most important role.
The two contributions are then summed in quadrature and one takes the $(\pm 1/2)$ 
of the square root to obtain a two-sided error bar.
Note that the finite-resolution uncertainties we quote are consistent 
with Table~II of~\cite{Bernuzzi:2011aq} and
provide an average between optimistic and conservative estimate of the
errors (obtained respectively by resolution-extrapolated data from
five and three simulations with different resolutions).  

From the top panel of Fig.~\ref{fig:freq_mod} one sees visually how
the tidal models clearly yield a better agreement with the numerical
data than the simple point-mass EOB model~\footnote{ %%
  Note that up to GW frequency $M\omega_{22}\sim 0.1$ the Taylor~T4
  point-mass phasing agrees very well with the NR one and thus with 
  the NR-tuned EOB point-mass.}. 
This information is made more quantitative in the the bottom panels,
where the point-mass analytical prediction is seen to deviate away
starting from $2000M$, while the other differences remaining much flatter,
and marginally close to the error bar, up to contact.  The comparison
of Fig.~\ref{fig:freq_mod} allows us  to deduce the presence of tidal
effects in the very late part of the inspiral, just before
contact. However, it also indicates that this comparison is sensible
{\it at most} to leading-order tidal effects, since, given the
uncertainties on the NR data, it is not possible to meaningfully
disentangle 1PN and 2PN tidal corrections from the leading-order ones.
On these plots, the Taylor T4 PN model with LO tidal corrections
looks consistent with the EOB predictions, yielding similar differences
with the NR waveform.
Note that at the very contact position the frequency and amplitude differences
show a clear increasing trend. This might either be due to the ``blurred'' 
nature of contact in the NR data, or to the lack of suitably determined 
next-to-quasi-circular corrections. For simplicity, we will not overtune
here more our analytical model and postpone to a future investigation the
detailed analysis of these additional effects.

We discuss now the phasing by means of a gauge-invariant and
frequency-based analysis employing the
$Q_\omega=\omega^2/\dot{\omega}$ function (where we put $\omega\equiv
M\omega_{22}$ for simplicity) introduced and used extensively in
Ref.~\cite{Baiotti:2011am}. We recall that the meaning of this
function is that the time-domain GW phase $\phi_{(\omega_1,\omega_2)}$
accumulated between frequencies $(\omega_1,\omega_2)$ is given by the
integral
$\phi_{(\omega_1,\omega_2)}=\int_{\omega_1}^{\omega_2}Q_\omega d\ln
\omega$.  Consequently, a change of $Q_\omega(\omega)$ of the order
$\pm 1$ during a frequency octave $\ln(\omega_2/\omega_1)$ corresponds
to a local dephasing (around $\omega$) of $\delta\phi\simeq \pm
1~rad$.  The main advantage of the $Q_\omega$ diagnostics is that it
is independent of the arbitrary time and phase shifts $(\tau,\alpha)$
necessary to compare the waveforms in the time domain.

As in Ref.~\cite{Baiotti:2011am} we cannot compute $Q_\omega$ from the
raw NR data, but we have fitted the phase of $h_{22}$ with a suitable
PN-based expression (see Eqs.~(27)-(28)
of~\cite{Baiotti:2011am}). Here the best fit is given by using a
sixth-order polynomial (in contrast to the fourth-order polynomial
employed in~\cite{Baiotti:2011am}) in the variable
$x=\left[\nu(t_m-t)/5\right]^{-1/8}$ (where $t_m$ is a fitting 
parameter formally representing the merger time), and the time interval
$[t_L,t_R]/M=[965,2400]$, which corresponds to frequencies
$[\omega_L,\omega_R]=[0.042,0.063]/M$.  
The $Q_\omega$ curve, so obtained, is represented by the 
thick-solid line with circles in Fig.~\ref{fig:Qomg}. The shaded 
region around the curve indicates the uncertainty on the curve as
given by $Q_\omega^{\rm NR}\pm \sigma_{ Q_\omega^{\rm NR}}$.
This numerical uncertainty $\sigma_{Q_\omega^{\rm NR}}$ is estimated by
putting together the effect of truncation error, of finite
extraction radius, and of the fit. To do so, we first calculated 
other three $Q_\omega$ curves,: one from the waveform of the M run, 
$Q_\omega^{\rm   NR_{\rm M}}$; another from the waveform of the H run, 
but extrapolated at infinite extraction radius, $Q_\omega^{\rm
  NR^\infty_H}$; a third one doing the fit of the H data with 
the fourth-order polynomial instead of the sixth-order one, 
$Q_\omega^{{\rm NR}_{{\rm H},n=4}}$. We then computed the 
differences $\delta Q^{{\rm H},\infty}_\omega=Q_\omega^{\rm NR_H}-Q_\omega^{\rm NR^\infty_H}$, 
$\delta Q^{\rm HM}_\omega=Q_\omega^{\rm NR_H}-Q_\omega^{\rm NR_{\rm M}}$ and
$\delta Q^{{\rm H},n=4}_\omega=Q_\omega^{\rm NR_H}-Q_\omega^{{\rm NR}_{{\rm H}, n=4}}$ and
summed them in quadrature, 
so to estimate the error-bar $\sigma_{Q_\omega^{\rm NR}}=\pm
1/2\sqrt{\left(\delta Q^{\rm HM}_\omega\right)^2 
+ \left(\delta Q^{{\rm H}\infty}_\omega\right)^2 
+ \left(\delta Q^{{\rm H},n=4}_\omega\right)^2 }$ 
that we represented in Fig.~\ref{fig:Qomg}. This conservative error
estimate is of order unity, as it varies between $(-3,+2)$ at
$\omega=0.0415$, and between $(-1,+1)$ at $\omega=0.0603$. 

Together with the numerical curve we also exhibit in the picture other
five analytical EOB curves: the point-mass EOB (dash-dotted, black
online), the LO tidal EOB (thick-dashed, blue online), the analytical, 
2PN tidal EOB with $\alpha_2^{(2)}=85/14$ (thick-solid, red online) and 
the effective 2PN tidal EOB with the {\it effective} values 
$\bar{\alpha}_2=40$ and $\bar{\alpha}_2=100$ used in Ref.~\cite{Baiotti:2011am} 
(lowermost dotted lines, black and magenta online, from top to bottom). 
The figure highlights clearly the dependence of NR on tidal interaction. 
One concludes that: (i) the NR curve is always very close to the 2PN tidal
EOB one; (ii) it is very well distinguishable from the point-mass prediction;
(iii) the error-bar on the NR curve is too large to
appreciate the differences between the LO and 2PN tidal EOB models;
(iv) the effective 2PN tidal EOB model used in Ref.~\cite{Baiotti:2011am} with
$\bar{\alpha}_2=100$ significantly overestimates the magnitude of tidal
interactions in the NR data; (v) the effective 2PN tidal EOB model 
with $\bar{\alpha}_2=40$ gives a good average of the numerical points.

The most important information suggested by
Fig.~\ref{fig:Qomg} is that the EOB tidal model constructed using {\it only}
analytically computed tidal information is by itself consistent
with the NR simulation, {\it without} the need of tuning any additional tidal EOB
flexibility parameter yielding an effective amplification of the tidal
interaction as the stars get closer and closer. We can not exclude
that such an amplification exists~\footnote{ 
  Especially on the basis of the analytical considerations of
  Ref.~\cite{Bini:2012gu} suggesting that $\alpha_{2}^{(2)}$ might be
  replaced by an effective distance dependent coefficient $\alpha_2^{\rm
    eff}(u)\equiv \alpha_2^{(2)}/(1-r_{\rm LR}u)$, $r_{\rm LR}$ denoting the
  EOB effective light-ring location. Note however that the corrisponding
  $Q_\omega$ curve would be indistingushable from the nonresummed one on 
  the plot.}, 
but one will need much higher accuracy in the late inspiral phase to
identify actual physical effects. 
In this respect, although the value $\bar{\alpha}_2=40$ fits well the
NR data, it does not indicate definitely an amplification of tidal
effects, as truncation errors are still dominant in this frequency
range and the value of $\bar{\alpha}_2$ is very sensitive 
to small changes on $Q_\omega$ that are barely visible on the plot.
For instance, note how the EOB $Q_\omega$ is easily matching the upper 
bound of the error-bar by taking $\bar{\alpha}_2=20$. We expect that the use of
higher-resolutions and/or more accurate numerical treatments of the  
hydrodynamics will further move up the NR $Q_\omega$ curve, so 
to favour smaller values of $\bar{\alpha}_2$ than larger ones.

The (visually) small differences between the
$Q_\omega$'s in Fig.~\ref{fig:Qomg} actually correspond to relevant
dephasings, of order $1~rad$ or more. This information, relative to
the frequency interval $[\omega_1,\omega_2]M=[0.041,0.062]$, is
quantified in Table~\ref{tab:dphi}. 
The 2PN (NNLO) tidal EOB model accumulates a dephasing of $1.06~rad$,
the LO EOB model $1.49~rad$, while the point-mass EOB $3.92~rad$. 
The uncertainties on these numbers are of the order
$\sigma_{\Delta\phi}=0.61~rad$, and obtained by integrating 
the shaded region in Fig.~\ref{fig:Qomg}. 

Due to the fitting procedure involved in the computation of the
$Q_\omega$ curves~\cite{Baiotti:2011am}, it is important to verify the
phasing with another diagnostic. Hence, we present also an analysis
based on waveform alignment as customary in the literature.
The time evolution of the phase difference
$\Delta\phi_{22}(t)=\phi^{\rm X}(t)-\phi^{\rm NR}(t)$ (where the label X can
be either EOB or T4) is shown in Fig.~\ref{fig:phasing}. It is computed 
from the time-and-phase alignement waveforms, as in Fig.~\ref{fig:freq_mod}.  
The shaded region represents the uncertainty on the NR phase.
The qualitative information given by this plot confirms the analysis 
of the phasing given by the $Q_\omega$: tidal effects are clearly visible
before contact and the current analytical knowledge is sufficient to
match the NR phasing up to contact (dashed vertical line in the figure). 
It is not possible, however, to distinguish in the NR data the effect
of higher-order tidal effects from LO ones.  
Interestingly, on this plot the T4 tidal LO model performs marginally 
worse than the corresponding EOB model, $\sim -0.3~rad$ at contact
and notably out of the estimated NR uncertainty.
Note that the phase difference  varies in the range 
$\pm 0.2~rad$ from the beginning of the simulation 
(after the initial burst) up to contact. This value 
is consistent with lower value  
$\Delta\phi-\sigma_{\Delta\phi}\sim 0.5$ obtained from Table~\ref{tab:dphi}.
For completeness, in Fig.~\ref{fig:phasing} we also added the two phase
differences with the effective 2PN tidal EOB model, with $\bar{\alpha}_2=20$ 
and $\bar{\alpha}_2=40$. The are both well within the error bar, although, 
consistently with the $Q_\omega$ analysis, the value $\bar{\alpha}_2=20$ 
actually yields a smaller dephasing at contact.
In conclusion, putting together the information of Figs.~\ref{fig:Qomg}-\ref{fig:phasing},
state-of-the art numerical simulations allow us to conclude that, if any
actual amplification of tidal effects exists, it yields, conservatively, 
$\bar{\alpha}<40$ (as a conservative estimate), or, more likely, $\bar{\alpha}_2<20$.

We conclude this Section with a few comments about the accuracy of the
higher-order multipoles, that are actually included in the computation
of $E^{\rm NR}(j)$. Figure~\ref{fig:l3l4} compares the modulus of the
most relevant subdominant numerical multipoles $\ell=3$, $m=2$ and
$\ell=m=4$ with the corresponding EOB waveforms. To our knowledge,
this comparison has never been shown before. The visual agreement is rather 
good, with both analytical multipoles averaging the corresponding numerical 
ones practically up to contact, as it was the case for the $\ell=m=2$ case. 
From the picture one also sees the large initial burst of junk radiation that 
must be included in the accurate computation of the $E^{\rm NR}(j)$ relation.

\ifusesec
\section{Conclusions}
\else
\paragraph*{Conclusions}
\fi
\label{sec:conc}

In this paper we have presented a comparison between dynamics and
waveform from BNS coalescence computed from long-term ($\sim$ ten orbits) 
NR simulations and the tidal EOB model including all the known tidal PN
corrections~\cite{Bini:2012gu}.

New numerical simulations have been presented which improve
quantitatively previous results~\cite{Bernuzzi:2011aq}. A set of
simulations which employ the same initial data, grid setup, and
resolutions of~\cite{Bernuzzi:2011aq}, but adopt an higher-order
reconstruction method in the HRSC scheme and time integrator, 
has shown that tidal effects can be overestimated by numerical inaccuracies. 
While the data show convergent behavior before the contact, around
this point and later the uncertainties related to numerical viscosity
do not seem completely under control, and actually become dominant
over truncation errors.

In order to compare NR data with the analytical EOB model, we estimated the
GW frequency of the contact as, $M\omega^{\rm c}_{22}\sim 0.078$. 

The dynamics of the system has been investigated by means of the 
$E(j)$ relation between the reduced binding energy $E$ and the 
reduced angular momentum $j$, computed here for the first time for 
BNS simulations and presented in Fig.~\ref{fig:E_Vs_j}. The 
tidal EOB model is consistent with the NR data up to contact ($j=3.63$) 
and even later, up to $j\sim3.5$. 

The effects of tidal interactions are clearly visible in the NR/EOB
waveforms. The comparison of amplitude and frequency
(Fig.~\ref{fig:freq_mod}) indicates that tidal effect become significant in
the very late part of the inspiral, just before contact. Given the
uncertainties on the NR data, it is not possible to meaningfully
disentangle 1PN (NLO) and 2PN (NNLO) tidal corrections from the LO
ones.  The T4 tidal model with LO tidal corrections is slightly 
worse (-0.3 rad) than the EOB model when getting close to contact.
The phasing was studied by means of both a gauge-invariant,
frequency-based analysis employing the $Q_\omega$
diagnostic~\cite{Baiotti:2011am} and a standard time and phase-shift 
alignment procedure. The results of the former method are collected in
Fig.~\ref{fig:Qomg} and Table~\ref{tab:dphi}, the ones of the latter
in Fig.~\ref{fig:phasing}. The $Q_\omega$ diagnostic is more affected
by the noise of the data which results in somehow larger
uncertainties; the time and phase alignment suffers of ambiguities in
the choice of the interval and potentially  underestimates actual
differences accumulated up to the alignment interval. In summary, while they
give slightly different numbers, the picture emerging from the two
phasing analyses is consistent: tidal effects are clearly visible
during the late inspiral up to contact and the current analytical 
knowledge is sufficient to match the NR phasing. It is not possible, 
however, to distinguish in the NR data the effect of higher-order 
tidal effects from LO ones. We observe that after contact and up 
to merger (i.e., for about one further orbit), the 2PN EOB model 
performs better than any other analytical tidal model; note however
that the extension of any analytical model beyond contact has
only an effective meaning.

In conclusion, we have shown that the current analytic knowledge
incorporated into the EOB model is sufficient to reproduce within 
the uncertainties the numerical data up to
contact. No calibration of any tidal effective-one-body free
parameter is required, beside those already fitted to binary black
holes data. While the 2PN (NNLO) model minimizes the differences with
the NR data, it is not possible to significantly distinguish it from
the 1PN (LO) model. Obviously, we cannot exclude the presence of a further
amplification of tidal interaction as the star gets close 
(as suggested by Ref.~\cite{Bini:2012gu}), but the present NR data 
indicate that this effect, if present, is {\it smaller} than 
what believed in  the past~\cite{Baiotti:2011am} (i.e. $\bar{\alpha_2}=100$) 
and it is not possible to estimate it precisely. A conservative analysis
points points to $\bar{\alpha}_2<40$, though we think that a more 
likely estimate (at one-sigma level) is $\bar{\alpha}_2<20$.
Note in addition that for higher, more realistic compactnesses 
(say ${\cal C}=0.16-0.18$) tidal effects are even smaller, thus
potentially more difficult to extract from the numerical data.
Similar considerations also hold for the use of realistic EOS,
which present their own numerical challanges to be used in NR
simulations.

As a consequence, the 2PN-accurate tidal EOB model~\cite{Bini:2012gu} 
used in this work~\footnote{Here we focused on the equal-mass case 
only, but we expect this EOB model to be accurate in the range of plausible mass ratios
for a BNS system, from $M_A/M_B=1$ to $M_A/M_B=0.7$, corresponding to $\nu\in[0.2422,0.25]$.
The reason for this being the fact that the EOB model used here was found consistent
with (now relatively old) BBH data up to 2:1 mass ratio ($\nu=2/9=0.2222$).
Further improvements of the current EOB point-mass model are currently 
in progress~\cite{DNRP}, consistently with the finding of Ref.~\cite{Pan:2011gk}.}
should be considered in the future as the most reliable choice to produce 
exact/target data for the development of templates for data-analysis 
purposes~\cite{Damour:2012yf}. 

This work also pointed out the importance of extensive numerical tests
to assess the uncertainties of the numerical data, and the potential 
need of new numerical strategies to perform accurate simulations.
Considering that the simulations presented here are the longest and
employ the among the highest resolutions to date, error assessment and
convergence tests appear absolutely necessary in future studies of this
kind. Because the use of significantly higher
resolutions (e.g.~$\sim400^3$ points covering each star) and extensive
tests seem to be computationally unfeasible, the development of
alternative and more accurate numerical methods seems unvoidable to
further improve and confirm our results.

\begin{acknowledgments}
  We warmly thank T.~Damour for prompting this research and for a careful reading of the manuscript.
  This work was supported in part by  
  DFG grant SFB/Transregio~7 ``Gravitational Wave Astronomy''.
  SB thanks IHES for hospitality during the development of part of this work.
  Computations were performed on JuRoPA (JSC), SuperMUC (LRZ), and
  Louhi (CSC) clusters. Computer time in Louhi was granted by PRACE Tier-1.
\end{acknowledgments}

\bibliography{refs20120515}{}

\end{document}